\documentclass{aastex631}

\usepackage{soul}

\begin{document}

\title{SN 2021hpr: A Normal Type Ia Supernova Showing Excess Emission in the Early Rising Phase}

\author[0009-0003-9229-9942]{Abdusamatjan Iskandar}
\affiliation{Xinjiang Astronomical Observatory, Chinese Academy of Sciences, Urumqi, Xinjiang, 830011, China}
\affiliation{School of Astronomy and Space Science, University of Chinese Academy of Sciences, Beijing 100049,  China}


\author[0000-0002-7334-2357]{Xiaofeng Wang} 
\correspondingauthor{Xiaofeng Wang}
\email{wang\_xf@mail.tsinghua.edu.cn}
\affiliation{Physics Department, Tsinghua University, Beijing, 100084, China}

\author{Ali Esamdin} 
\correspondingauthor{Ali Esamdin}
\email{aliyi@xao.ac.cn}
\affiliation{Xinjiang Astronomical Observatory, Chinese Academy of Sciences, Urumqi, Xinjiang, 830011, China}
\affiliation{School of Astronomy and Space Science, University of Chinese Academy of Sciences, Beijing 100049,  China}

\author{Xiangyun Zeng} 
\affiliation{Center for Astronomy and Space Sciences, Three Gorges University, YiChang, 443000, People's Republic of China}
\affiliation{College of Sciences, Three Gorges University, YiChang, 443000, China}


\author{Craig Pellegrino} 
\affiliation{Goddard Space Flight Center, 8800 Greenbelt Rd, Greenbelt, MD 20771, USA}

\author{Shengyu Yan} 
\affiliation{Physics Department, Tsinghua University, Beijing, 100084, China}

\author{Jialian Liu} 
\affiliation{Physics Department, Tsinghua University, Beijing, 100084, China}

\author{Alexei V. Filippenko} 
\affiliation{Department of Astronomy, University of California, Berkeley, CA 94720-3411, USA}

\author{D. Andrew Howell} 
\affiliation{Las Cumbres Observatory, 6740 Cortona Drive Suite 102, Goleta, CA 93117-5575, USA}
\affiliation{Department of Physics, University of California, Santa Barbara, CA 93106-9530, USA}

\author{Curtis McCully} 
\affiliation{Las Cumbres Observatory, 6740 Cortona Drive Suite 102, Goleta, CA 93117-5575, USA}
\affiliation{Department of Physics, University of California, Santa Barbara, CA 93106-9530, USA}

\author{Thomas G. Brink} 
\affiliation{Department of Astronomy, University of California, Berkeley, CA 94720-3411, USA}

\author{Maokai Hu} 
\affiliation{Physics Department, Tsinghua University, Beijing, 100084, China}

\author{Yi Yang} 
\affiliation{Department of Astronomy, University of California, Berkeley, CA 94720-3411, USA}
\affiliation{Physics Department, Tsinghua University, Beijing, 100084, China}

\author{WeiKang Zheng} 
\affiliation{Department of Astronomy, University of California, Berkeley, CA 94720-3411, USA}


\author{Guoliang L$\ddot{\rm u}$} 
\affiliation{Xinjiang Astronomical Observatory, Chinese Academy of Sciences, Urumqi, Xinjiang, 830011, China}

\author{Jujia Zhang}  
\affiliation{Yunnan Observatories (YNAO), Chinese Academy of Sciences, Kunming, 650216, China}
\affiliation{International Centre of Supernovae, Yunnan Key Laboratory, Kunming, 650216, China}

\author{CuiYing Song} 
\affiliation{Physics Department, Tsinghua University, Beijing, 100084, China}

\author{RuiFeng Huang} 
\affiliation{Physics Department, Tsinghua University, Beijing, 100084, China}

\author{Rachael Amaro} 
\affiliation{Department of Astronomy and Steward Observatory, The University of Arizona, 933 North Cherry Avenue, Tucson, AZ 85721, USA}

\author{Chunhai Bai}  
\affiliation{Xinjiang Astronomical Observatory, Chinese Academy of Sciences, Urumqi, Xinjiang, 830011, China}

\author{Kyle G. Dettman} 
\affiliation{Department of Physics and Astronomy, Rutgers, the State University of New Jersey, Piscataway, NJ 08854,USA}

\author[0000-0002-1296-6887]{Llu\'is Galbany} 
\affiliation{Institute of Space Sciences (ICE-CSIC), Campus UAB, Carrer de Can Magrans, s/n, E-08193 Barcelona, Spain}
\affiliation{Institut d'Estudis Espacials de Catalunya (IEEC), 08860 Castelldefels (Barcelona), Spain}

\author[0000-0002-1125-9187]{Daichi Hiramatsu} 
\affiliation{Center for Astrophysics \textbar{} Harvard $\&$ Smithsonian, 60 Garden Street, Cambridge, MA 02138-1516, USA} \affiliation{The NSF AI Institute for Artificial Intelligence and Fundamental Interactions, USA}
\author{Bostroem K. Azalee} 
\affiliation{Steward Observatory, University of Arizona, 933 North Cherry Avenue, Tucson, AZ 85721, USA}

\author{Koichi Itagaki} 
\affiliation{Itagaki Astronomical Observatory, Yamagata, Yamagata 990-2492, Japan}

\author[0000-0001-8738-6011]{Saurabh W. Jha} 
\affiliation{Department of Physics and Astronomy, Rutgers, the State University of New Jersey, Piscataway, NJ 08854,USA}

\author{Shuguo Ma} 
\affiliation{Key Laboratory of Optical Astronomy, National Astronomical Observatories, Chinese Academy of Sciences, Beijing, 100012, China}

\author{David Sand} 
\affiliation{Steward Observatory, University of Arizona, 933 North Cherry Avenue, Tucson, AZ 85721-0065, USA}

\author[0000-0003-0123-0062]{Jennifer E. Andrews}
\affiliation{Gemini Observatory, 670 North A`ohoku Place, Hilo, HI 96720-2700, USA}

\author[0000-0001-5754-4007]{Jacob Jencson} 
\affiliation{IPAC, Mail Code 100-22, Caltech, 1200 E. California Blvd., Pasadena, CA 91125, USA}

\author[0000-0002-0370-157X]{Peter Milne} 
\affiliation{Steward Observatory, University of Arizona, 933 North Cherry Avenue, Tucson, AZ 85721-0065, USA}

\author[0000-0002-0744-0047]{Jeniveve Pearson} 
\affiliation{Steward Observatory, University of Arizona, 933 North Cherry Avenue, Tucson, AZ 85721-0065, USA}

\author[0000-0002-4022-1874]{Manisha Shrestha} 
\affiliation{Steward Observatory, University of Arizona, 933 North Cherry Avenue, Tucson, AZ 85721-0065, USA}

\author[0000-0001-5510-2424]{Nathan Smith}
\affiliation{Steward Observatory, University of Arizona, 933 North Cherry Avenue, Tucson, AZ 85721-0065, USA}

\author[0000-0002-8770-6764]{R\'eka K\"onyves-T\'oth}  
\affiliation{Konkoly Observatory, HUN-REN Research Center for Astronomy and Earth Sciences, Konkoly
Th. M. út 15-17., Budapest, 1121 Hungary; MTA Center of Excellence}
\affiliation{Department of Experimental Physics, Institute of Physics, University of Szeged, D\'om t\'er 9, Szeged, 6720 Hungary}

\author{Samuel Wyatt} 
\affiliation{Astrophysics Science Division, NASA Goddard Space Flight Center, Greenbelt, MD 20771, USA}

\author{Xuan Zhang} 
\affiliation{Xinjiang Astronomical Observatory, Chinese Academy of Sciences, Urumqi, Xinjiang, 830011, China}






\author{Shahidin Yaqup} 
\affiliation{Xinjiang Astronomical Observatory, Chinese Academy of Sciences, Urumqi, Xinjiang, 830011, China}





\author{Letian Wang} 
\affiliation{Xinjiang Astronomical Observatory, Chinese Academy of Sciences, Urumqi, Xinjiang, 830011, China}

\author{Mengfan Zhang} 
\affiliation{Xinjiang Astronomical Observatory, Chinese Academy of Sciences, Urumqi, Xinjiang, 830011, China}

\def\dm15{1.00 $\pm$ 0.01}

\begin{abstract}
We present extensive optical observations of a nearby Type Ia supernova (SN Ia), SN 2021hpr, located in the spiral galaxy NGC 3147 at a distance of $\sim$ 45 Mpc. Our observations cover a phase within $\sim$ 1--2 days to $\sim 290$ days after the explosion. SN 2021hpr is found to be a spectroscopically normal SN Ia, with an absolute \textit{B}-band peak magnitude of $M_{\rm max}(B) \approx -19.16 \pm 0.14$ mag and a post-peak decline rate of $\Delta m_{15}(B)=$ \dm15 mag. ‌Early-time light curves showed a $\sim 7.0 \%$ excess emission compared to a homogeneously expanding fireball model, likely due to SN ejecta interacting with a companion or immediate circumstellar matter. The optical spectra of SN 2021hpr are overall similar to those of normal SNe Ia, but characterized by prominent detached high-velocity features (HVFs) of Si {\sc ii} and Ca {\sc ii} in the early phase. After examining a small sample of well-observed normal SNe Ia, we find that the HVFs are likely common for the subgroup with early-excess emission. The association of early bump feature with the HVFs could be attributed to density or abundance enhancement at the outer layer of the exploding star, likely as a result of interactions with companion/CSM or experiencing more complete burning. Nevertheless, the redshifted Fe {\sc ii} and Ni {\sc ii} lines in the nebular-phase spectra of SN 2021hpr, contrary to the blueshift trend seen in other SNe Ia showing early bump features, indicate its peculiarity in the explosion that remains to be understood.

\end{abstract}



\section{Introduction} \label{sec:intro}
Type Ia supernovae (SNe Ia) have been widely used as cosmological distance indicators because of their relatively high and uniform peak luminosities \citep[e.g.,][]{Phillips1993ApJ,Riess1996ApJ,Wang2005ApJ,Burns2018ApJ}. Observations of SNe Ia in the local and distant universe have led to the discovery of the accelerating expansion of the universe \citep{Riess1998AJ,Perlmutter1999ApJ}. 
Despite the widespread belief that SNe~Ia stem from the thermonuclear explosions of carbon-oxygen white dwarfs (WDs), significant debates persist over the progenitor system and the explosive processes \citep{Woosley1986ApJ,Nugent2011Natur,Bloom2012,Wang2012NewAR,Maoz2014,Darnley2014A&A,Jha2019NatAs}.
The observed characteristics of SNe Ia exhibit growing diversity in both photometric and spectroscopic measurements over time, and subclassifying them can enhance the precision of distance measurements \citep{Taubenberger2017hsn,Wang2009ApJ,Wang2013Sci}.

The nature of the donor star to the WD remains unclear, with any luminous red giant companion being excluded for the well-known nearby object SN 2011fe \citep{Li2011Natur}.
Two popular models for progenitor systems include the single-degenerate (SD) scenario, where the CO WD accretes material from a nondegenerate companion star and is triggered to produce a thermonuclear explosion when its mass is close to the Chandrasekhar mass limit \citep{Chandrasekhar1957,Whelan1973,Nomoto1982,Wang2009MNRAS},  and the double-degenerate (DD) scenario, where the companion star is another WD --- either the dynamic merger or collision of two WDs triggers a runaway thermonuclear explosion \citep{Iben1984,Webbink1984,Pakmor2012,Kushnir2013ApJ}. 
In the SD case, hydrogen- or helium-rich circumstellar matter (CSM) is expected to exist around the progenitor system. Interactions of SN Ia ejecta with CSM have been reported for a few objects as support for the SD scenario \citep{Dilday2012Sci,Maguire2013MNRAS,Silverman2013ApJS,Hu2023MNRAS}; however, the absence of hydrogen features in the nebular spectra of SNe~Ia is still a challenge for the SD model \citep{Maguire2013MNRAS,Maguire2016MNRAS,Silverman2013ApJS,Tucker2020MNRAS,Lim2023}. In the DD case, a few explosion mechanisms have been proposed, including dynamical merger (steady accretion from the secondary WD), double detonation (the detonation of He on the primary WD surface triggers a carbon detonation in its core), violent merger of two WDs, and a head-on collision as the detonation could be triggered directly by shock ignition, rather than the propagation and acceleration of any subsonic deflagration burning front \citep{Pakmor2010Natur,Woosley1986ApJ,Bildsten2007ApJ,Kushnir2013ApJ}. In dynamical mergers, if the accretion rate is high enough to ignite carbon, then the WD may collapse to a neutron star instead of becoming an SN Ia explosion. However, the exact explosion physics and progenitor systems of SNe Ia are still unclear \citep{Wang2012NewAR,Wang2013Sci,Maoz2014,Jha2019NatAs,Taubenberger2017hsn,Tucker2020MNRAS,Maeda2022hxga}. 

Observationally, $\sim70 \%$ of SNe Ia can be classified as spectroscopically normal SNe Ia \citep{Branch1993AJ,Li2011MNRAS}, while the remaining 30\%  can be categorized into different kinds of peculiar subclasses, such as over-luminous SN 1991T-like, subluminous SN 1991bg-like, and low-luminosity SN 2002cx-like SNe Ia \citep{Filippenko1992AJ,Leibundgut1993AJ,Filippenko1992ApJ,Filippenko1997,Li2003PASP,Foley2013ApJ}.
According to the velocity gradient of Si {\sc ii} $\lambda 6355$, \cite{Benetti2005ApJ} divided SNe Ia into three subclasses: the high-velocity gradient (HVG), the low-velocity gradient (LVG), and the FAINT SNe Ia, which are similar to the SN 1991bg-like. Based on the minimum velocity measured from Si {\sc ii} $\lambda 6355$ absorption at around the \textit{B}-band maximum light, \cite{Wang2009ApJ} classified normal SNe Ia into two subclasses: normal-velocity (NV) and high-velocity (HV). This classification reveals that even Branch-normal SNe Ia should have different progenitor populations \citep{Wang2013Sci}. 

Very early-time photometric and spectroscopic observations of SNe Ia provide additional important constraints on their progenitor properties. Theoretically, \cite{Kasen2010ApJ} predicted that in the SD scenario, the SN ejecta should run into the nondegenerate donor star and be heated by the shock. The shock-heated material would produce detectable optical/ultraviolet (UV) emission lasting for hours to days after the explosion, depending on the size of the donor star, the pre-explosion binary separation, the viewing angle, and the expansion velocity of the ejecta. 
On the other hand, double detonation of a carbon-oxygen (CO) WD and/or mixing of radioactive $^{56}$Ni into the outer region of the ejecta have been proposed to explain the "bump" in the early light curves \citep{Noebauer2017MNRAS.472.2787N,Jiang2017Natur.550...80J,Polin2019ApJ,Magee2020}.

In recent years, some wide-field and high-cadence surveys have led to the discovery of many young SNe Ia, among which a few samples are reported to show early excess emission in their light curves such as SNe 2013dy, 2019np, 2018oh, 2017cbv, 2017erp, 2020hvf, 2021aefx, 2023bee \citep{Zheng2013ApJ,Pan2015MNRAS,Sai2022MNRAS,Burke2022,Levanon2019ApJ,Li2019ApJ,Shappee2019ApJ,Dimitriadis2019ApJ,Hosseinzadeh2017ApJ,Wang2020ApJ,Jiang2021ApJ,Ashall2022ApJ,Ni2023arXiv,Hosseinzadeh2023ApJ,Wang2024ApJ}. Based on observation simulations, the intrinsic fraction of SNe Ia with early flux excesses is estimated as $28^{+13}_{-11}$\% \citep{Magee2022MNRAS}. 

\cite{Burke2022} fit the early-time light curves of 9 SNe Ia using the companion interaction model and suggested that three of them (SN 2017cbv, SN 2017erp, and SN 2018yu) exhibit early excesses that may result from this model, while the best-fitting parameters of SN 2019np are unusual and therefore do not confidently claim an excess. In \cite{Sai2022MNRAS}, the early-time light curves of SN 2019np were reported to show an excess that can be attributed to the mixing of radioactive $^{56}$Ni\citep{Sai2022MNRAS}. Although the early bumps seen in the light curves of SN 2017cbv can be explained by collision of the SN ejecta with the main-sequence companion, late-time spectral analysis does not favor the presence of an H- and/or He-rich secondary star in the progenitor system \citep{Sand2018ApJ,Hosseinzadeh2017ApJ}. 
Despite the detection of an early bump in the light curve of SN 2018oh, the deep mixing of carbon in its ejecta and the lack of an H line in the nebular spectra cannot be well explained by the current SD model \citep{Li2019ApJ,Tucker2019ApJ,Dimitriadis2019ApJ,Graham2022MNRAS}. \cite{Maguire2016MNRAS} present the tentative detection of $\rm H_{\alpha}$ emission for SN 2013ct in the late-time spectra, but the estimated mass ($\sim$ 0.007 $\rm M_{\odot}$ ) of the stripped companion star material is much lower than expected in SD scenarios.
\cite{Dimitriadis2019ApJ} suggested that such an early bump could be caused by the interaction of SN ejecta with a disk formed during the merger process of a WD binary system. SN 2021aefx is another SN Ia showing bump features in early optical and UV light curves, but none of current models can account for the early excess emission \citep{Hosseinzadeh2022ApJ}. On the other hand, some peculiar subluminous SNe Ia like iPTF 14atg \citep{Cao2015Natur} and SN 2022vqz \citep{Xi2024MNRAS} are reported to show much stronger excess emission at early times relative to that seen in normal SNe Ia, which may have a different physical origin (i.e., double detonation of a subchandrasekhar-mass CO WD). 

SN 2021hpr is another nearby SN Ia with very early detection, providing another opportunity to constrain the progenitor physics of SN Ia from early-time luminosity evolution. 
Although SN 2021hpr has been studied by \cite{Zhang2022PASPhpr} and \cite{Lim2023}, we present more extensive photometric and spectroscopic observations in this paper, allowing us to conduct a more thorough analysis of its progenitor properties. The optical observations and data reduction are given in Section \ref{sec:obs}. Section \ref{Section:LV} describes the evolution of light and color curves, while the optical spectra are presented in Section \ref{Section:Optspectra}. The quasibolometric light curve and late-time spectra are discussed in Section \ref{Section:discuss}, and we summarize in Section \ref{section:conclusion}.

\section{OBSERVATIONS AND DATA REDUCTION} \label{sec:obs}

\subsection{Discovery and Host Galaxy}
\begin{figure}[ht!]
        \centering
	\includegraphics[width=4in]{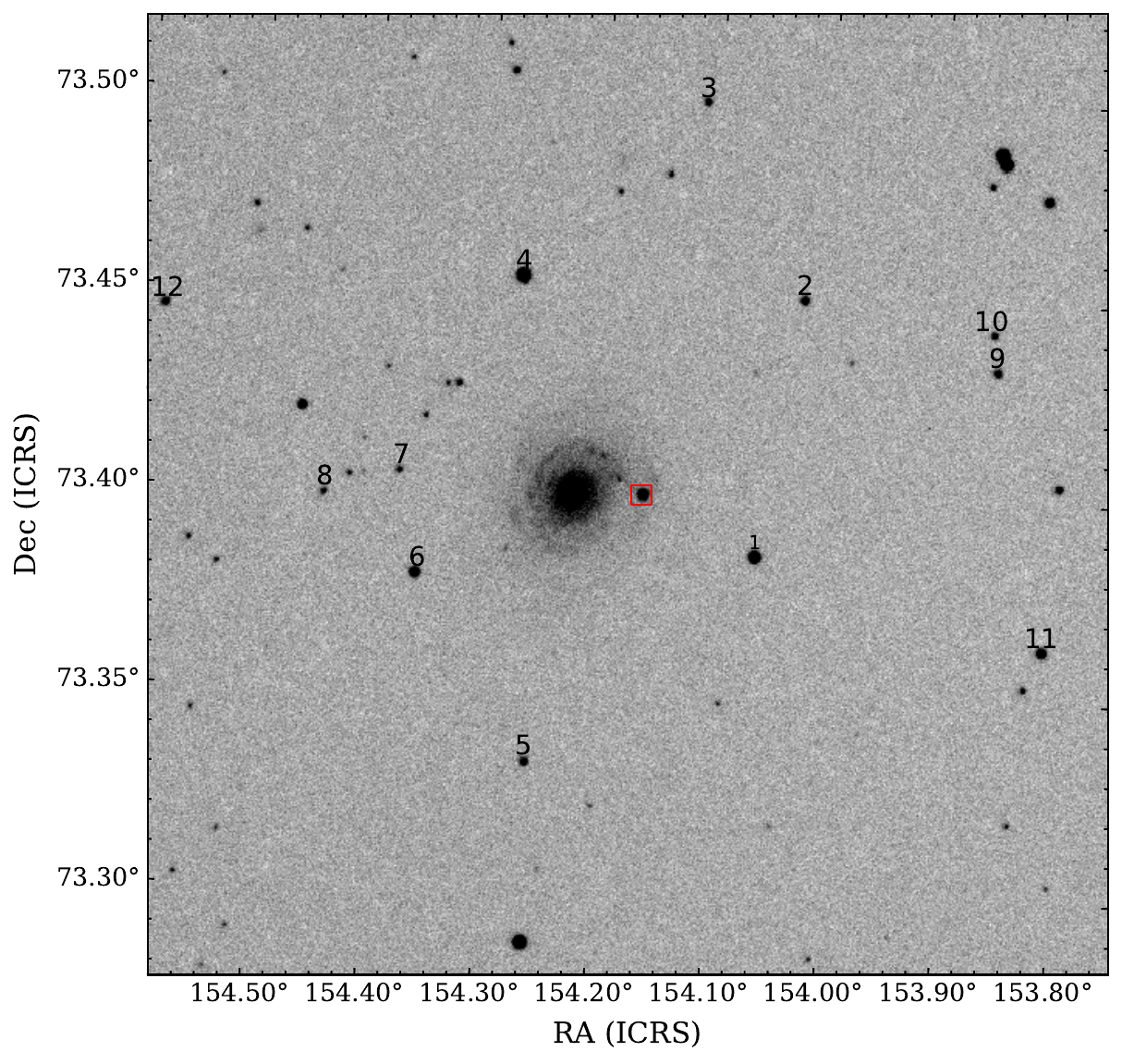}
	\caption{ \textit{B}-band image of SN 2021hpr taken with the NOWT on 2021 Apr. 27.88. The red square represents the SN, while numbers indicate the reference stars used for photometry. North is up and east is to the left.}
	\label{fig:finder}
\end{figure}

SN 2021hpr was discovered by Koichi Itagaki on 2021 April 2.4489 (MJD  59,306.4489; UTC dates are used throughout this paper) with an unfiltered brightness of 17.7 mag \citep{Itagaki2021}. Its J2000 coordinates are $ \alpha =  10^{\rm hr}16^{\rm m}38.68^{\rm s}$ and $\delta = +73\degr 24'01.80''$, located $\sim 224.5''$ east and $\sim 1.1''$ north of the center of the barred spiral galaxy NGC 3147. Prediscovery detection can be traced back to 2021 March 31.92 (MJD 59,304.92), about 1.52 days earlier than the discovery date, obtained with the RC600 60\,cm telescope of the Caucasian Mountain Observatory at $B \approx 19.35$ mag \citep{Lim2023}.  
This SN was classified as a young SN Ia according to a spectrum taken by the Asiago Ekar Telescope \citep{Tomasella2021} at $\sim 1.07$ days after the discovery. Figure \ref{fig:finder} shows a finder chart of SN 2021hpr and its host galaxy. 

\startlongtable
\begin{deluxetable*}{lccccccccccc}
\tablecolumns{6} 
\tablewidth{0pc} 
\tabletypesize{\scriptsize}
\tablecaption{Local reference stars in the SN 2021hpr field from the APASS catalog$^a$}
\tablehead{\colhead{Star} &\colhead{$\alpha$ (J2000)} &\colhead{$\delta$ (J2000)}  &\colhead{$B$ (mag)} &\colhead{$V$ (mag)} &\colhead{$g$ (mag)} &\colhead{$r$ (mag)} &\colhead{$i$ (mag)} }
\startdata
		1&154.06203&73.38566 & 13.903(026)&13.554(066)&13.832(191)&13.505(140)&13.472(193)\\
		2&154.02423&73.45025 & 15.634(065)&14.920(070)&15.397(181)&14.688(125)&14.472(160)\\
		3&154.11455&73.49906 & 16.091(023)&15.295(068)&15.779(174)&15.097(103)&14.810(169)\\
		4&154.27259&73.45459 & 13.268(029)&12.544(069)&13.012(180)&12.351(128)&12.185(173)\\
		5&154.25859&73.33306 & 15.709(084)&15.051(090)&15.489(187)&14.855(129)&14.583(166)\\
		6&154.35740&73.37931 & 14.815(023)&14.132(078)&14.568(180)&13.952(123)&13.816(191)\\
		7&154.37454&73.40489 & 16.627(125)&15.965(027)&16.448(163)&15.814(183)&15.631(161)\\
		8&154.44091&73.39899 & 16.761(114)&16.061(014)&16.563(203)&15.819(091)&15.668(145)\\
        9&153.85333&73.43334 & 15.709(049)&14.748(093)&15.354(178)&14.368(140)&14.054(173)\\
        10&153.85693&73.44261& 16.127(103)&15.638(052)&15.952(225)&15.457(124)&15.245(146)\\
        11&153.80875&73.36343& 14.690(046)&14.129(071)&14.524(186)&13.971(122)&13.863(200)\\
		12&154.58534&73.44488& 15.702(028)&15.273(096)&15.587(182)&15.216(131)&15.032(047)\\
\enddata
\tablenotetext{}{$^a$Uncertainties are in units of 0.001 mag. See Figure \ref{fig:finder} for the locations of reference stars.}
\label{tab:Sstar}
\end{deluxetable*}

The host galaxy NGC 3147 is an SA(rs)bc galaxy \citep{Vaucouleurs1991rc3}, at redshift $z = 0.009346$ \citep{Epinat2008MNRAS}, which harbored six known SN explosions over the past $\sim 50$\,yr, including SN 1972H (SN Ia, \citealt{Patat1997}), SN 1997bq (SN Ia, \citealt{Jha2006AJ}), SN 2006gi (SN Ib, \citealt{Duszanowicz2006}), SN 2008fv (SN Ia, \citealt{Biscardi2012}), SN 2021do (SN Ic, \citealt{Voziakova2021}), and SN 2021hpr. The birth rate of SNe Ia seems to be unusually high in NGC 3147 \citep{Biscardi2012, Zhang2022PASPhpr}. 

\subsection{Photometry} 
Our follow-up observations of SN 2021hpr were conducted with a few telescopes, including the Las Cumbres Observatory (LCO) 1\,m telescope network \citep{Shporer2011IAUS,Brown2013PASP},
the 0.76\,m Katzman Automatic Imaging Telescope (KAIT) as part of the Lick Observatory Supernova Search \citep[LOSS;][]{Filippenko2001},
and the Nanshan One-meter Wide-field Telescope (NOWT) at Nanshan Station of the Xinjiang Astronomical Observatory (XAO) \citep{Bai2020RAA}. Some of the data are from the Global Supernova Project.
The NOWT is equipped with a 4k $\times$ 4k pixel CCD camera, with a field of view that covers $1.6\degr \times 1.6\degr$. Its observations of SN 2021hpr were conducted in the \textit{UBV} bands during the period from $\sim 5$ days to $\sim 53$ days after the explosion. The images were processed with the IRAF package\footnote{IRAF, the Image Reduction and Analysis Facility, is distributed by the National Optical Astronomy Observatories, which are operated by the Association of Universities for Research in Astronomy (AURA), Inc., under a cooperative agreement with the National Science Foundation.}, which includes bias subtraction and flat-field correction. 
The instrumental magnitudes were determined using the SExtractor software \citep{Bertin1996}.

KAIT obtained $BVRI$-band images, as well as $Clear$ (close to the $R$ band; see \citealt{Li2003}) images. All images were reduced using a custom pipeline\footnote{https://github.com/benstahl92/LOSSPhotPypeline} detailed by \citet[][]{Stahl2019}. Point spread function (PSF) photometry was obtained using {\tt DAOPHOT} \citep[][]{Stetson1987} from the {\tt IDL} Astronomy User’s Library\footnote{http://idlastro.gsfc.nasa.gov/}. Several nearby stars were chosen from the Pan-STARRS1\footnote{http://archive.stsci.edu/panstarrs/search.php} catalog for calibration; their magnitudes were first transformed into Landolt \citep{Clem2016AJ} magnitudes using the empirical prescription presented by Equation 6 of \citet[][]{tonry_pan-starrs1_2012}, and 
then transformed to the KAIT natural system. The apparent magnitudes were all measured in the KAIT4 natural system, and the final results were transformed into the standard system using the local calibrator and the color terms for KAIT4 \citep[see][]{Stahl2019}.

LCO observations sampled the \textit{UBVgri} bands, covering the phases from $\sim 1$ day to $\sim 80$ days. 
To reduce the images, we utilize both \textit{lcogtsnpipe} \citep{Valenti2016MNRAS} and a PyRAF-based pipeline that includes bias subtraction, flat-field correction, and SN flux measurement. 
The instrumental magnitudes obtained by LCO and NOWT are calibrated using the AAVSO Photometric All-Sky Survey (APASS; https://www.aavso.org/apass) catalog \citep{Henden2016JAVSO}. 
The local reference stars with photometric magnitudes from the APASS catalog are listed in Table \ref{tab:Sstar} and labeled in Figure \ref{fig:finder}. 
The \textit{U}-band instrumental magnitudes were converted to those of the standard Johnson \textit{UBV} system, based on the transformation correlations established through Landolt standard stars observed on photometric nights using NOWT \citep{Clem2016AJ}. 

This SN was also monitored by the 0.35\,m telescope of Itagaki Astronomical Observatory (IAO)  in the \textit{Clear} band, covering the phases from $\sim 2$ days to $\sim 14$ days.
Photometry was extracted using \texttt{Astrometrica} \citep{Raab2012ascl} and calibrated to the 4th CCD Astrograph Catalog 4th of the US Naval Observatory (UCAC4; \citealt{Zacharias2013AJ}).

No template subtraction is applied to the NOWT, KAIT, LCO, and IAO images in the photometry, because the SN locates relatively far away from the center of its host galaxy. In addition, we included the ZTF \textit{gr}-band data\footnote{http://134.158.75.151:24000/ZTF21aarqkes.} as well as the ATLAS cyan- and orange-band data \citep{Tonry2018ApJ} in our analysis, covering the phases from $\sim 1$ to $\sim 266$ days.

\subsection{Spectroscopy}
A total of 33 low-resolution optical spectra of SN 2021hpr, covering the phases from $\sim -15$ days to $\sim +288$ days after \textit{B}-band maximum light, were collected using the LCO~2\,m Faulkes Telescope North \cite[FTN;][]{Brown2013PASP}, the YFOSC on the Lijiang 2.4\,m telescope \cite[LJT;][]{Wang2019RAALJT} of Yunnan Astronomical Observatories, the BFOSC mounted on the Xinglong 2.16\,m telescope \cite[XLT;][]{Jiang1999xlt,Fan2016PASPxlt,Zhang2016PASPxlt}, and the Kast spectrograph on the Lick 3\,m Shane telescope \citep{Stahl2020MNRAS}. Another nebular phase spectrum was obtained on Jan. 31 2022 with the Low-Resolution Imaging Spectrometer (LRIS) mounted on the Keck 10~m telescope \citep{Oke1995PASP}. See Table \ref{tab:speclog} for the journal of spectral observations. We used the standard IRAF routines and performed flux calibration with spectrophotometric standard stars observed with similar air masses. We further applied atmospheric extinction corrections and telluric absorption corrections to all spectra. We also included nine public spectra from the Weizmann Interactive Supernova Data Repository (WISeREP) in the analysis, including six from XLT (previously published in \citealp{Zhang2022PASPhpr}) and three posted on Transient Name Server, obtained with the Schmidt-Cassegrain Telescope (SCT, Meade 10$''$), the ALPY 200 telescope at Three Hills Observatory (THO), and the Liverpool Telescope (LT).

\label{Section:spec}
\startlongtable
\begin{deluxetable*}{lccccc}
\tablecolumns{6} 
\tablewidth{0pc} 
\tabletypesize{\scriptsize}
\tablecaption{Log of Spectroscopic Observations of SN 2021hpr}
\tablehead{\colhead{MJD} & \colhead{Telescope/Instrument} & \colhead{Epoch$^a$} & \colhead{$\lambda_{\rm start}-\lambda_{\rm end}$} & \colhead{$\Delta\lambda$}\\
           \colhead{days} &                                &            \colhead{days}    &        \colhead{$\mathring{\rm A}$}  &\colhead{$\mathring{\rm A}$}}
\startdata
          59,307.23       & LCO/FTN      & $-$14.88 & 3500.44-9999.81   & 2.29\\
          59,307.52       & XLT/BFOSC$^{\rm W}$    & $-$14.59 & 3701.35-8847.19   & 2.79\\
          59,307.93       & THO/ALPY200$^{\rm W}$  & $-$14.18 & 4004.12-7699.96   & 4.70\\
          59,308.01       & LT/SPRAT$^{\rm W}$     & $-$14.11 & 4020.00-7966.80   & 9.20 \\
          59,309.30       & Lick/Kast    & $-$12.81 & 3620.00-10,720.00  & 2.00\\
          59,310.50       & XLT/BFOSC    & $-$11.61 & 3863.83-8824.22   & 2.77\\
          59,311.50       & XLT/BFOSC    & $-$10.61 & 3859.70-8838.58   & 2.77\\
          59,313.50       & XLT/BFOSC    & $-$8.61  & 3844.53-8803.51   & 2.77\\
          59,314.36       & LCO/FTN      & $-$7.75  & 3499.47-9999.73   & 2.29\\
          59,316.50       & XLT/BFOSC    & $-$5.61  & 3856.20-8816.73   & 2.77\\
          59,317.23       & LCO/FTN      & $-$4.88  & 3500.37-9999.64   & 2.29\\
          59,317.58       & XLT/BFOSC$^{\rm W}$       & $-$4.53  & 3702.04-8844.84   & 2.72\\
          59,319.17       & Lick/Kast    & $-$3.94  & 3620.00-10,750.00  & 2.00\\
          59,320.23       & LCO/FTN      & $-$1.88  & 3499.93-9999.93   & 2.29\\
          59,322.16       & Lick/Kast    & +0.05    & 3632.00-10,660.00  & 2.00\\ 
          59,323.23       & LCO/FTN      & +1.12    & 3500.49-10,000.20  & 2.29\\
          59,323.50       & XLT/BFOSC    & +1.39    & 3861.65-8819.92   & 2.77\\
          59,323.92       & SCT Meade 10$''$/SN Spec$^{\rm W}$     & +1.81    & 3897.84-7152.43   & 1.36\\
          59,324.09       & LJT/YFOSC    & +1.98    & 3503.49-8766.44   & 2.83\\
          59,328.24       & LCO/FTN      & +6.13    & 3500.44-10,000.76  & 2.29\\
          59,329.50       & XLT/BFOSC    & +7.39    & 3869.36-8836.11   & 2.77\\
          59,333.50       & XLT/BFOSC    & +11.39   & 3869.36-8836.11   & 2.77\\
          59,335.50       & XLT/BFOSC$^{\rm W}$       & +13.39  & 3702.39-8859.43   & 2.72\\
          59,336.23       &MMT/Binospec  & +14.12   & 5206.00-7702.00   &0.37\\          
          59,336.50       & XLT/BFOSC    & +14.39   & 3869.36-8835.04   & 2.77\\
          59343.14       & MMT/Binospec  & +21.03   & 5688.00-7209.00   & 0.37\\  
          59,344.25       & LCO/FTN      & +22.14   & 3499.93-10,000.00  & 2.29\\
          59,344.54       & XLT/BFOSC$^{\rm W}$       & +22.43  & 3700.07-8859.75   & 2.72\\
          59347.13       & MMT/Binospec   & +25.02   & 5206.00-7702.00  & 0.37\\  
          59,352.50       & XLT/BFOSC    & +30.39   & 3875.51-8838.16   & 2.77\\
          59,363.29       & LCO/FTN      & +41.18   & 3500.39-9999.81   & 2.29\\
          59,368.58       & XLT/BFOSC$^{\rm W}$       & +46.47  & 3700.17-8860.31   & 2.72\\
          59,369.50       & XLT/BFOSC    & +47.39   & 3869.23-8834.72   & 2.77\\
          59,385.56       & XLT/BFOSC$^{\rm W}$       & +63.45  & 3700.37-8858.61   & 2.72\\
          59,585.32       & Lick/Kast    & +263.21  & 3622.00-10,380.00 & 2.00\\ 
          59,610.50       & Keck/LRIS    & +288.39  & 3166.31-10,265.62 & 0.62\\ 
\enddata
\tablenotetext{a}{Relative to the date of \textit{B}-band maximum brightness (MJD = 59,322.11).}
\tablenotetext{w}{Data from \textit{wiserep}.}
\label{tab:speclog}
\end{deluxetable*}

\begin{figure} [ht!]
        \centering
	\includegraphics[width=5.in]{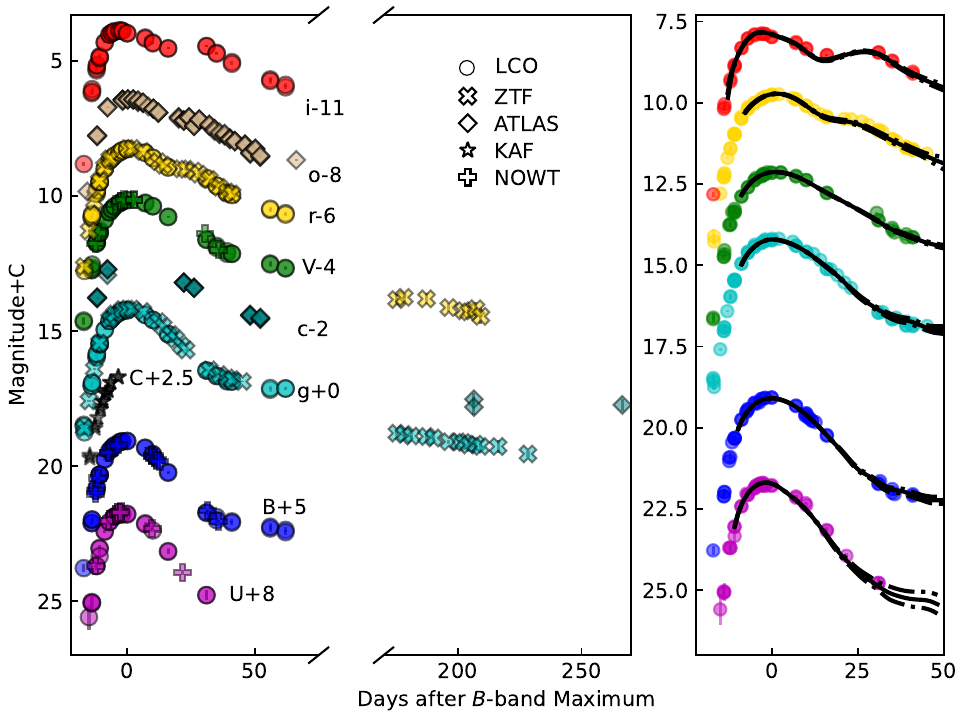}
	\caption{{\it Left panel:} Multiband light curves of SN 2021hpr, which are shifted vertically for better clarity. {\it Right panel:} Fitting to the observed light curves with the SNooPy2+ \texttt{Max model}. The dashed lines represent 1$\sigma$ uncertainties of the best fits.} 
	\label{fig:hpr_lc} 
\end{figure}

\section{Photometric Properties}
\label{Section:LV}
\subsection{Optical Light Curves}
The multiband optical light curves of SN 2021hpr are presented in the left panel of Figure \ref{fig:hpr_lc}, with a shoulder feature in \textit{r} and a secondary maximum in  \textit{i} like normal SNe~Ia.
Applying polynomial fits to the near-maximum light curves, we estimate the peak magnitudes and the corresponding time in all bands;  the relevant results are listed in Table \ref{tab:fit}. Around the maximum light, the \textit{B}-band light curve reached a peak of 14.11 $\pm$ 0.04\,mag on MJD = 59,322.11 $\pm$ 0.58, with a post-peak decline measured as $\Delta m_{15}(B)$ = \dm15\,mag within the first 15 days after the peak \citep{Phillips1999AJ}. 
The \textit{V}-band light curve reaches a peak of 14.12 $\pm$ 0.03\,mag on MJD = 59,322.98 $\pm$ 0.61, later than the \textit{B}-band peak by 0.87\,day.
We also use the SuperNovae in object-oriented Python \cite[SNooPy2;][]{Burns2011AJ,Burns2014ApJ} light-curve fitter to fit the light curves. The best-fit results are presented in the right panel of Figure \ref{fig:hpr_lc}. The color stretch parameter $s_{BV}$ \citep{Burns2014ApJ} is measured to be 1.02 $\pm$ 0.03. In summary, SN 2021hpr has a standard light-curve morphology in optical bands, and the parameters inferred from our observations agree well with those estimated by \cite{Zhang2022PASPhpr} and \cite{Lim2023}.

\startlongtable
\begin{deluxetable*}{lccccccccccc}
\tablecolumns{6} 
\tablewidth{0pc} 
\tabletypesize{\scriptsize}
\tablecaption{Peak magnitudes and corresponding time estimated from polynomial fits to the observed light curves of SN 2021hpr.}
\tablehead{\colhead{Filter} &\colhead{$U$} &\colhead{$B$} &\colhead{$V$} &\colhead{$g$} &\colhead{$r$} &\colhead{$i$} }
\startdata
		MJD (Days) &59,321.46 $\pm$ 0.76 &59,322.11 $\pm$ 0.58& 59,322.98 $\pm$ 0.61&59,323.38 $\pm$ 0.11 &59,322.72 $\pm$ 0.39 &59,319.81 $\pm$ 0.20\\
		Peak(mag)&13.61 $\pm$ 0.04     &14.11 $\pm$ 0.04    & 14.12 $\pm$ 0.03    &14.20 $\pm$ 0.05    &14.24 $\pm$ 0.02    &14.83 $\pm$ 0.02\\
\enddata
\label{tab:fit}
\end{deluxetable*}

\startlongtable
\begin{deluxetable*}{lccccccccccc}
\tablecolumns{6} 
\tablewidth{0pc} 
\tabletypesize{\scriptsize}
\tablecaption{Comparison SNe with SN 2021hpr}
\tablehead{\colhead{SN} &\colhead{$\Delta m_{15}(B)$} &\colhead{Early Excess} &\colhead{References} }
\startdata
		SN 2021aefx&0.90 $\pm$ 0.02 &yes& \cite{Hosseinzadeh2022ApJ}\\
		SN 2023bee &0.75 $\pm$ 0.03 &yes& \cite{Hosseinzadeh2023ApJ} \\
  	SN 2017erp &1.05 $\pm$ 0.06 &yes& \cite{Brown2019ApJ,Burke2022}\\
		SN 2019np  &1.05 $\pm$ 0.04 &yes& \cite{Sai2022MNRAS} \\
  	SN 2017cbv &0.88 $\pm$ 0.07 &yes& \cite{Wee2018ApJ}\\
        SN 2013dy  &0.89 $\pm$ 0.01 &yes & \cite{Pan2015MNRAS,Zheng2013ApJ}\\
		SN 2011fe  &1.18 $\pm$ 0.08 &no & \cite{Zhang2016fe,Munari2013NewA}\\
		SN 2018gv  &0.96 $\pm$ 0.05 &no & \cite{Yang2020ApJ,Burke2022} \\
  	SN 2015F   &1.35 $\pm$ 0.03 &no & \cite{Cartier2017MNRAS}\\
		SN 2017hpa &1.02 $\pm$ 0.07 &no & \cite{Zeng2021ApJhpa} \\
\enddata
\label{tab:csn}
\end{deluxetable*}

\begin{figure*}[ht!]
        \centering
	\includegraphics[width=6.5in]{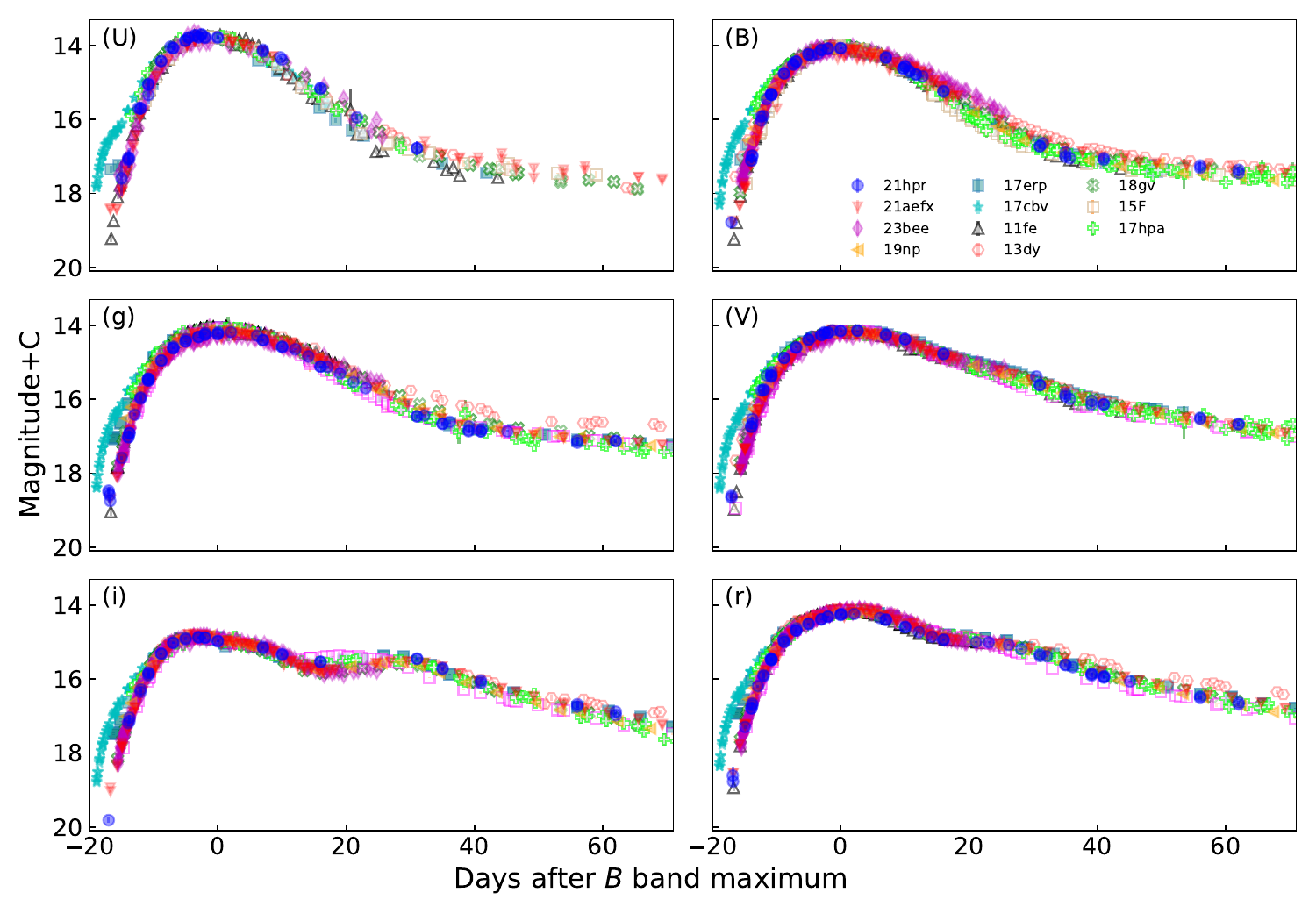}
	\caption{Comparison of the optical light curves of SN 2021hpr with those of other well-observed SNe Ia. The filled symbols represent SNe Ia with early excess emission, while the open symbols represent those without prominent excess emission. The light-curve peaks of comparison SNe Ia have been normalized to those of SN 2021hpr.}
	\label{fig:com_lc} 
\end{figure*}

In Figure \ref{fig:com_lc}, we compare the \textit{UBVgri}-band light curves of SN 2021hpr with those of well-observed normal SNe~Ia, including SN 2013dy, SN 2021aefx, SN 2023bee, SN 2017erp, SN 2019np, SN 2017cbv, SN 2011fe, SN 2018gv, SN 2015F, and SN 2017hpa (see Table \ref{tab:csn}).
The former six objects serve as representatives of those exhibiting early flux excess in the early phase after the explosion, whereas the latter represents those without early excess emission. The light-curve evolution of SN 2021hpr resembles that of those SNe~Ia displaying early flux excess, especially SN 2019np and SN 2021aefx. \cite{Hosseinzadeh2022ApJ} and \cite{Ni2023arXiv} suggested that SN 2021aefx shows a prominent red excess at early times. Section \ref{Section:5.3}, will discuss the early-excess features detected in SN 2021hpr.

\subsection{Reddening and Colors}	
\begin{figure*}[ht!]
        \centering
	\begin{minipage}[t]{0.45\linewidth}  
		\includegraphics[height=2.in,width=3.in]{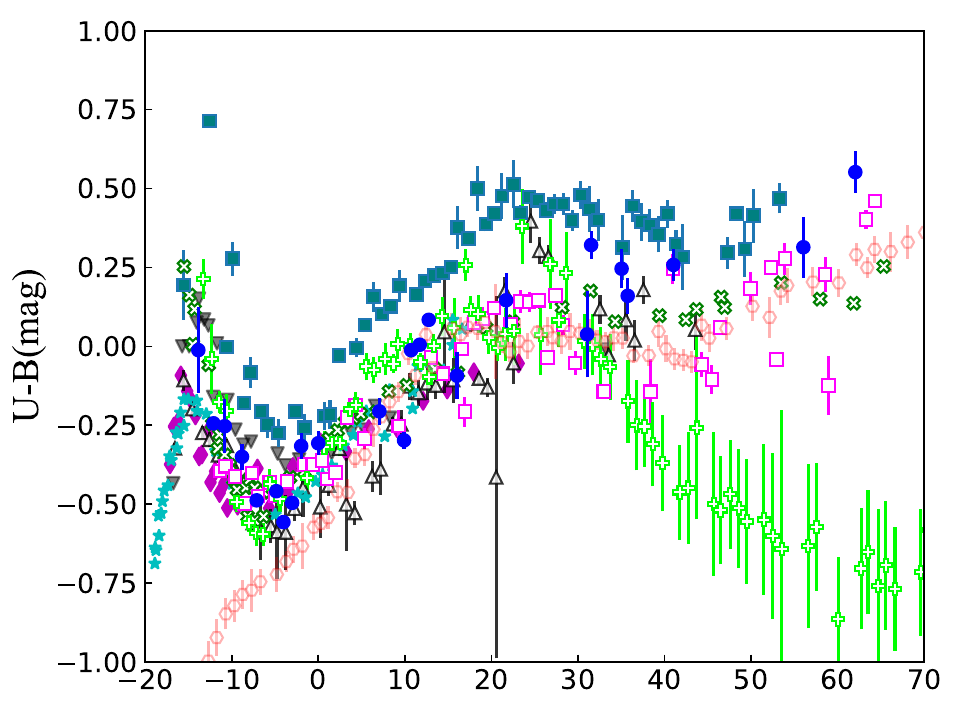}   
	\end{minipage}%
	\begin{minipage}[t]{0.45\linewidth}  
		\includegraphics[height=2.in,width=3.in]{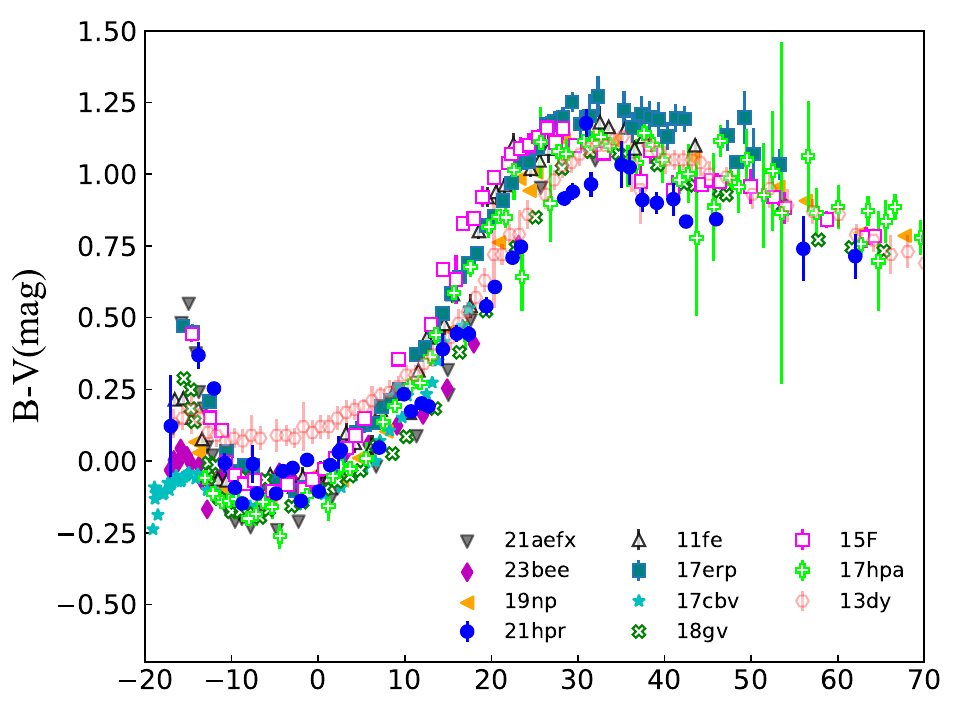}   
	\end{minipage} 

	\begin{minipage}[t]{0.45\linewidth}  
		\includegraphics[height=2.in,width=3.in]{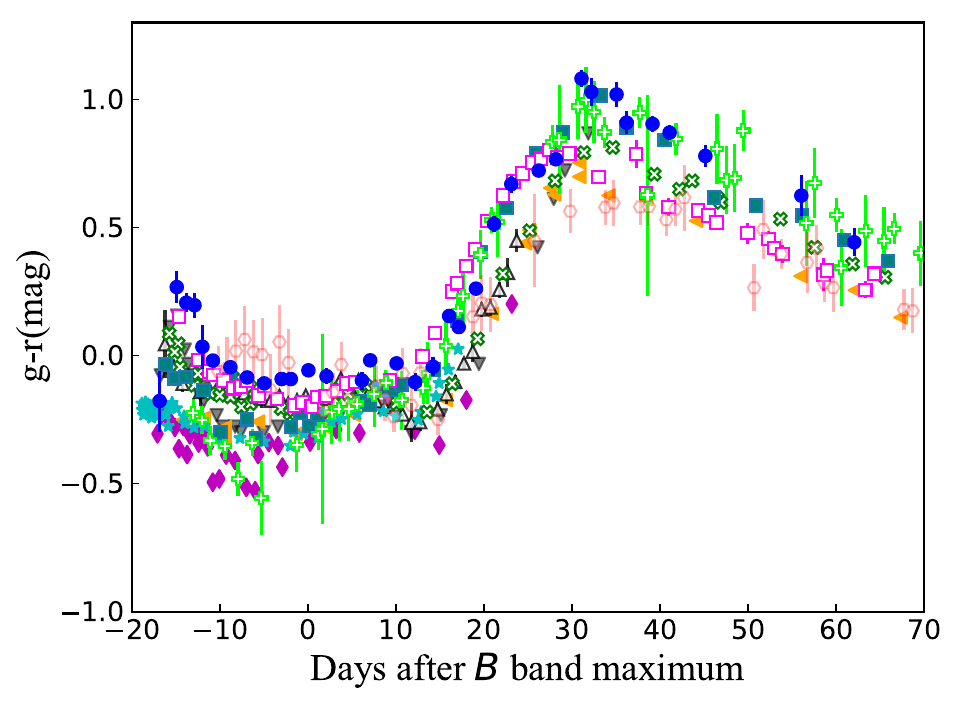}    
	\end{minipage}
	\begin{minipage}[t]{0.45\linewidth}  
		\includegraphics[height=2.in,width=3.in]{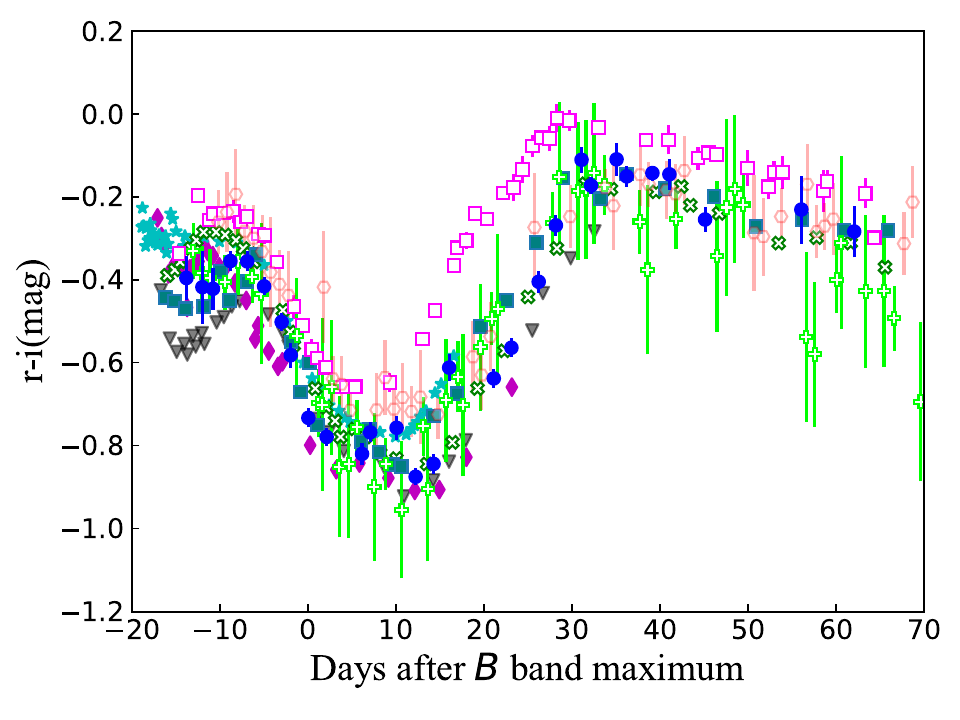}    
	\end{minipage}  
	\caption{Comparison of the reddening-corrected $U-B$, $B-V$, $g-r$, and $r-i$ color curves of SN 2021hpr and a few well-studied SNe Ia.  The comparison sample are the same as in Figure \ref{fig:com_lc}.
 }
	\label{fig:color}
\end{figure*}

The Galactic reddening towards the line of sight of SN 2021hpr is $E(B-V) \approx 0.021$\,mag \citep{Schlafly2011ApJ}. The host-galaxy reddening inferred from the \texttt{EBV model} of SNooPy2 gives $E(B - V )_{\rm host} = 0.06 \pm 0.06$\,mag. Based on the intrinsic $B-V$ color distribution of normal SNe~Ia \citep{Wang2009ApJ,Phillips1999AJ}, the $E(B - V )_{\rm host}$ is estimated as $0.054 \pm 0.015$\,mag. A lower reddening is also consistent with the observation that SN 2021hpr is located far from the arms and disk of NGC 3147. Thus, a total $E(B - V )_{\rm \rm total} = 0.08 \pm 0.06$\,mag is taken for SN 2021hpr. This value agrees well with the estimate from \cite{Lim2023}, $E(B - V )_{\rm total} = 0.10$\,mag. Figure \ref{fig:color} shows the reddening-corrected color curves of SN 2021hpr, overplotted with those of some well-observed objects (same as in Figure \ref{fig:com_lc}).

\subsection{First-Light Time}
\label{sec:FLT}
\begin{figure}[ht!]
	\centering
	\includegraphics[height=3.in,width=3in]{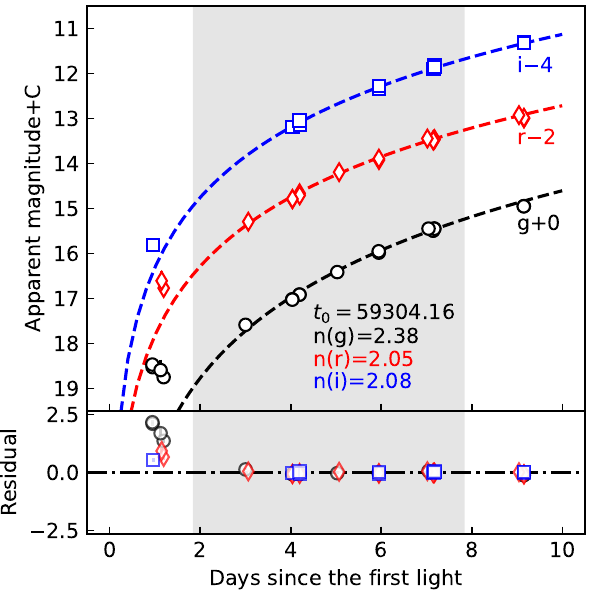}
	\caption{{\it Upper panel:} Fitting to the early-phase light curves of SN 2021hpr using the fireball model. {\it Bottom panel:} the residuals relative to the best-fit curves. The gray area represents the data used for the fit, ranging from t$\sim$ 2 days to 8 days after explosion.}
	\label{rise_time} 
\end{figure}

The very early light curves can be used to constrain the first-light time and the rise time. The prediscovery detection of SN 2021hpr at MJD 59,304.92 reported by \cite{Tsvetkov2021} suggests that the explosion time of SN 2021hpr should be earlier than this epoch. The last nondetection from ZTF is MJD 59,303.3452, with an upper limit of 19.77\,mag in the \textit{r} band. 
To estimate the first light time of SN 2021hpr, an expanding fireball model ($f\propto(t-\rm t_0)^n$; \cite{Riess1999AJ}) is adopted to fit the early-time \textit{gri} band light curves, as shown in Figure \ref{rise_time}. Considering the early excess of SN 2021hpr, we fit the data from t$\sim 2$ to t$\sim 8$ days after the explosion. The first light time is derived as MJD $59, 304.16 \pm 0.97$ days, with best fit index n = 2.38, 2.05, and 2.08 in \textit{g-, r-, and i} band, respectively. This indicates that our first observation began at t$\sim 1.1$ days after the SN explosion. As a result, the rise time is estimated as $17.95 \pm 1.13$ days in \textit{B} for SN 2021hpr, comparable to that of typical SNe Ia \citep{Zheng2017ApJ}. 

\section{Optical Spectra}
\label{Section:Optspectra}

\begin{figure} [ht!]
	\centering
	\includegraphics[width=4.8in]{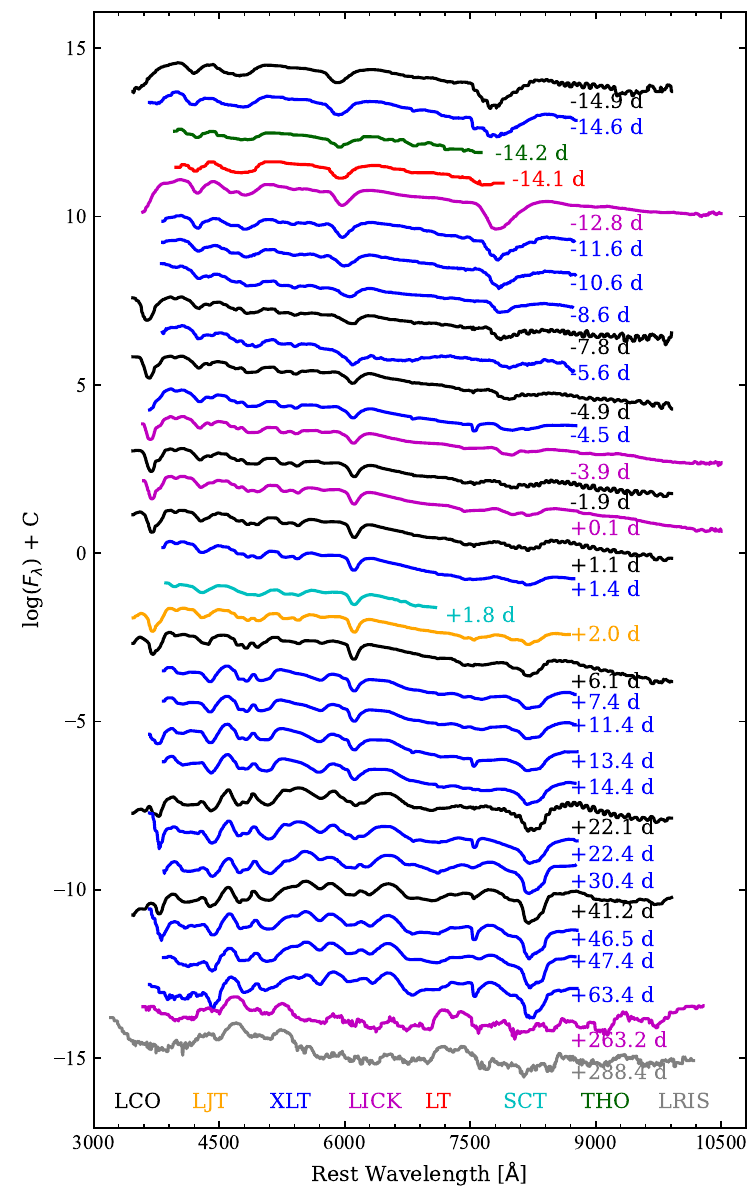}
	\caption{Optical spectral evolution of SN 2021hpr. All spectra have been corrected for reddening and host-galaxy redshift. The epochs shown on the right of the spectra represent the phases in days relative to the \textit{B}-band maximum light (MJD 59,322.11). The colors of the spectra represent data from different instruments.}
	\label{sepc1} 
\end{figure}

The spectral evolution sequence of SN 2021hpr, with nearly daily sampling before the maximum light, is shown in Figure \ref{sepc1}. Strong absorption features are clearly visible near $\lambda 5950$ and $\lambda 8150$ in the early phase but they become weak and redshifted quickly a few days later, which can be due to the presence of high-velocity features (HVFs) of the Si {\sc ii} and Ca {\sc ii} lines, respectively. Both carbon and oxygen absorption features are invisible or barely detectable in the spectra of SN 2021hpr, suggesting that the progenitor should have experienced a more complete burning during the explosion. The detailed spectral evolution will be discussed in the following subsections.


\subsection{Temporal Evolution}

\begin{figure*} [ht!]
	\centering
	\includegraphics[width=5.in,height=5.in]{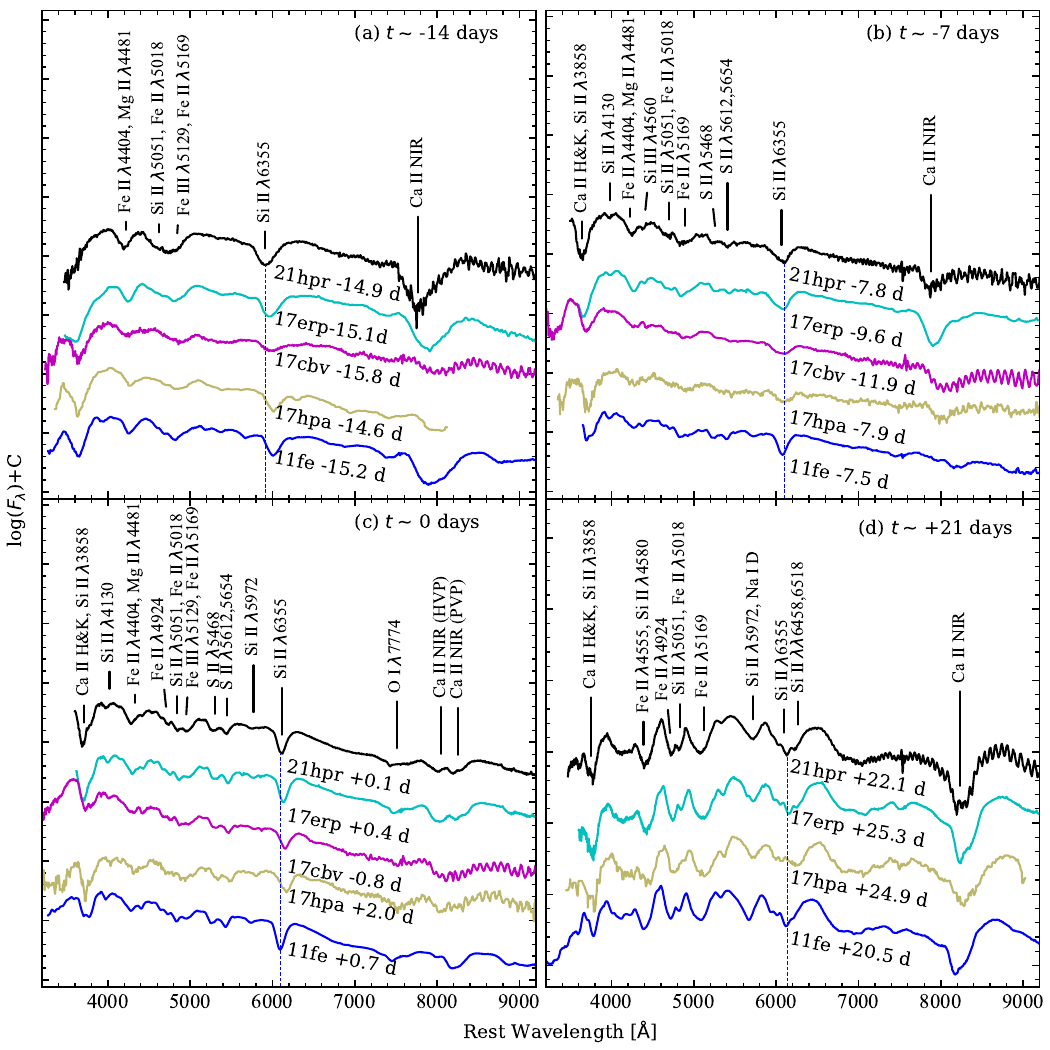}
	\caption{Spectral evolution of SN 2021hpr and the comparison SNe Ia at $t \approx -14$, $-7$, 0, and +21 days relative to the \textit{B}-band peak. All these spectra are shifted vertically for better clarity. The vertical dashed blue line marks the Si~{\sc ii} 6355\,\AA\ absorption minima in the spectra of SN 2021hpr.}
	\label{fig:sepc2} 
\end{figure*}
In Figure \ref{fig:sepc2}, we compare the spectra of SN 2021hpr with those of other well-observed SNe Ia with similar $\Delta m_{15}(B)$ at $t \approx -14$, $-7$, 0, and +21 days from the maximum light.
At t$\sim -14$ days, the main spectral features are overall similar to those of the comparison SNe~Ia (see Figure \ref{fig:sepc2}a), while SN 2021hpr has relatively broader absorption near $\sim 4600\,\mathring{\rm A}$ and larger ejecta velocity at this very early phase. 
The C~{\sc ii} $\lambda$6580 absorption feature is not visible in SN 2021hpr, while it is prominent in SN 2017erp and SN 2017hpa. 
At this early phase, the HVFs of Ca~{\sc ii} NIR triplet are prominent in SN 2021hpr, SN 2017erp, and SN 2011fe, while they are weaker in SN 2017hpa and SN 2017cbv. The HVF of Si~{\sc ii} $\lambda 6355$ absorption is also clearly seen in SN 2021hpr, while it is much less significant in the comparison objects, as shown in Figure \ref{fig:sepc2}a.

At $\sim 1$ weeks before maximum light (Figure \ref{fig:sepc2}b), the spectral features of SN 2021hpr becomes more similar to those of the comparison SNe~Ia. For example, the Si~{\sc ii} $\lambda 4130$ absorption begins to appear in the spectrum and the "W"-shaped S~{\sc ii} absorption feature develops near $\lambda 5400$. The blended absorption features near $\sim \lambda 4200$ and  $\sim \lambda 4600$ become separated. In comparison, the Ca~{\sc ii} NIR triplet still shows relatively large differences, as indicated by the relative strength of the HVFs and photospheric components among the comparison sample. For example, the Ca II HVFs are still strong in SNe 2021hpr, 2017erp, and SN 2017cbv while they tend to disappear in SN 2017hpa and SN 2011fe.  

Around the maximum light, the spectrum of SN 2021hpr shows close resemblances to that of the comparison SNe~Ia (see Figure \ref{fig:sepc2}c). At this phase, the line-strength ratio $R(\rm Si\,II)$ \citep{Nugent1995ApJ}, defined as the equivalent width (EW) ratio between Si~{\sc ii} $\lambda 5972$ and Si~{\sc ii} $\rm \lambda 6355$ in the near-maximum-light spectrum, is measured as $0.08 \pm 0.01$ in SN 2021hpr. 
This indicates that the photospheric temperature of SN 2021hpr is relatively higher.

By $t \approx 3$ weeks, the main spectral features of Ca~{\sc ii} H\&K, Si~{\sc ii}, iron-group elements, and even the Ca~{\sc ii} NIR triplet (which shows the most diversity at early phases) become similar for SN 2021hpr and the comparison sample.

\subsection{Photospheric Expansion Velocity}

\begin{figure*}[ht!]
        \centering
	\begin{minipage}[t]{0.45\linewidth}  
		\includegraphics[height=2.in,width=3.in]{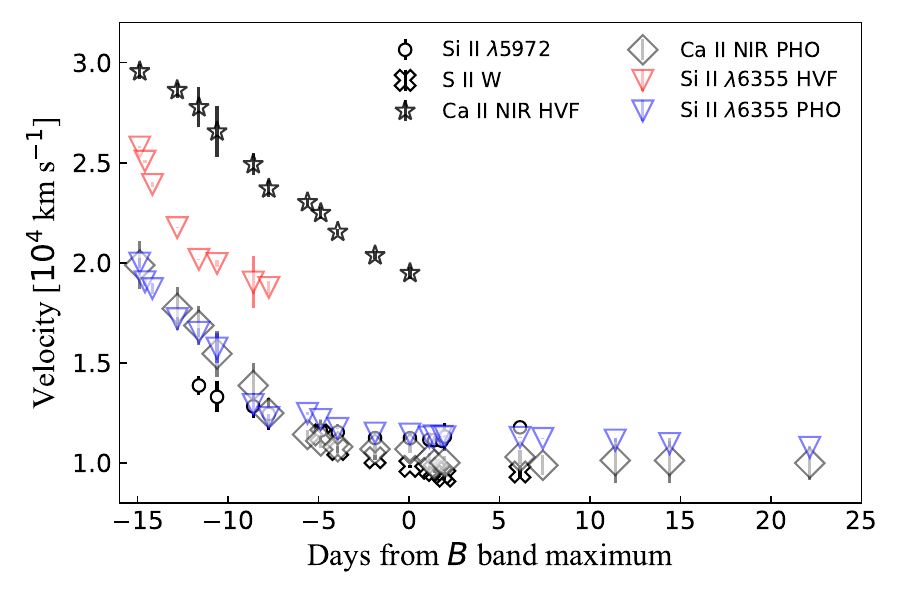}   
	\end{minipage}%
	\begin{minipage}[t]{0.45\linewidth}  
		\includegraphics[height=2.in,width=3.in]{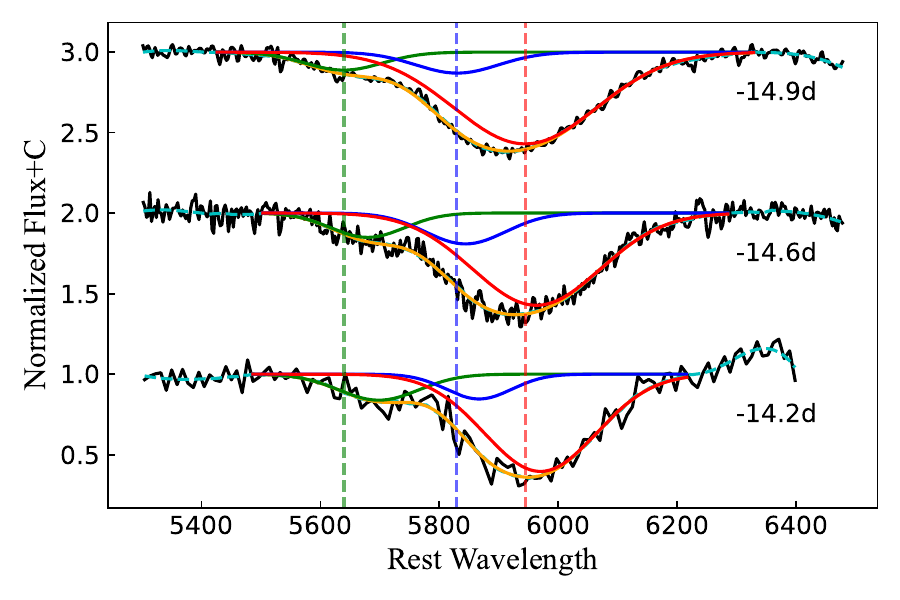}   
	\end{minipage} 
	\caption{{\it Left panel:} Expansion velocity of SN 2021hpr measured from absorption minima of Si~{\sc ii} $\lambda5972$, the "W"-shaped S~{\sc ii}, and the HVF and photospheric (PHO) components of the Si~{\sc ii} $\lambda6355$, and Ca~{\sc ii} NIR triplet. {\it Right panel:} The evolution of Si~{\sc ii} $\lambda 6355$ in SN 2021hpr compared with the multi-Gaussian fit at $\sim$ $-$ 14 days. Blue solid lines show the component on the blue side, and red solid lines show the component on the red side. The orange solid lines represent the best-fit curve for the observed profile, while the cyan dashed line represents the smoothed spectra. The color vertical lines represent the component minimum of each Gaussian function.}

	\label{fig:ovelocity}
\end{figure*}

Spectroscopic observations of SNe Ia provide a good opportunity to probe the layered structure of the photosphere. 
The left panel of Figure \ref{fig:ovelocity} shows the evolution of the expansion velocity of SN 2021hpr measured from the absorption minimum of the Si~{\sc ii} $\lambda 6355$ line. Measurement of the velocity of the spectral lines of Si~{\sc ii} $\lambda 5972$, "W"-shaped S~{\sc ii} was achieved by fitting a single Gaussian function to the absorption trough of the respective lines. For the single-gaussian fit, we employed the Monte Carlo random sampling method to derive an average error, which approximately corresponds to one standard deviation of the fit. The velocity of Si~{\sc ii} $\rm \lambda 6355$, Ca {\sc ii} NIR triplet in the projectile is calculated by applying a multi-Gaussian fit to the corresponding absorption lines in the spectrum \citep{Wang2009ApJcf,Zhao2015ApJS,Zhao2016ApJ}. In the first spectrum, the absorption lines of Si~{\sc ii} $\lambda$6355 and Si~{\sc ii} $\lambda$5972 are fitted with three-dimensional Gaussian functions to achieve better fitting results (see the right panel of Figure \ref{fig:ovelocity}). Considering that the absorption line of Si~{\sc ii} $\lambda$5972 will affect the double Gaussian fit on the absorption line of Si~{\sc ii} $\lambda$6355, three Gaussian functions are used for these two lines. The solid green lines represent the profile of the Si~{\sc ii} $\lambda$5972. The absorption component on the left (blue line) is considered as the high-velocity component of Si~{\sc ii} $\lambda$6355, with an estimated velocity of $\sim 25900\,\rm km\,s^{-1}$ at t$-14.9$ days. This velocity is about $\sim 6000\,\rm km\,s^{-1}$ larger than the photospheric velocity (red line), but about $\sim 4000\,\rm km\,s^{-1}$ smaller than the velocity of the Ca~{\sc ii} NIR HVF. This indicates that the HVF of Si~{\sc ii} $\lambda$6355 is less pronounced than that of Ca~{\sc ii} NIR.

At $t \approx -14.9$ days, the velocity inferred from Si~{\sc ii} $\lambda$6355 absorption is $20,000 \pm 258$\,\,km\,s$^{-1}$, which is similar to  that of SN 2017erp ($\sim 20,000\,\rm km\,s^{-1}$ at $\sim -16$ days; \citealt{Brown2019ApJ}), and much higher than that of SN 2011fe ($\sim 15,900\,\rm km\,s^{-1}$ at t = $-16.0$ days; \citealt{Zhang2016fe}). 
At $t \approx 0$ days, the Si~{\sc ii} velocity of SN 2021hpr drops to $11,453 \pm 100\,\rm km\,s^{-1}$. The velocity gradient derived during the phase from -14.9 to maximum light is thus $571 \pm 18\,\rm km\,s^{-1}\, day^{-1}$, which is close to that of SN 2017erp. Such a large velocity gradient suggests that SN 2021hpr may have undergone an asymmetric explosion  
 \citep{Maeda2010ApJ} or an interaction of ejecta with the circumstellar materials (CSMs) \citep{Gerardy2004ApJ}. 

Assuming that the velocity decreases in a parabolic way, we can estimate the expansion velocity of SN 2021hpr as  $\sim 11,500\,\rm km\,s^{-1}$ at around the maximum light, which is consistent with that measured from the $t \approx 0.1$ day spectrum. According to the method of \citet{Wang2009ApJ}, SN 2021hpr can be classified as a NV SN~Ia. Following the definition by \citep{Benetti2005ApJ}, we also measured the velocity gradient during the phase from $t \approx 1.1$ days to $t \approx 11.4$ days, giving a velocity gradient of $18 \pm 6$\,km\,s$^{-1}$\,day$^{-1}$. This puts SN 2021hpr into the LVG subclass of SNe~Ia. 


The left panel of Figure \ref{fig:ovelocity} presents the ejecta velocities measured from the absorption minima of ``W"-shaped S~{\sc ii}, Si~{\sc ii} $\lambda 5972$, Si~{\sc ii} $\lambda 6355$, and Ca~{\sc ii} NIR triplet. During the phase from $t \approx -14.9$ to $-5.6$ days, the strong HVFs dominate the Ca~{\sc ii} NIR triplet. The photospheric component starts to become pronounced thereafter. Following the method proposed by \citet{Zhao2015ApJS,Zhao2016ApJ}, the velocity of the Ca~{\sc ii} NIR HVF is measured as $\sim 29,500\,\rm km\,s^{-1}$ at $t \approx -14.9$ days, much higher than the corresponding Si~ {\sc ii} $\lambda6355$ velocity, while the photospheric component of the Ca~ {\sc ii} NIR triplet appears to have a velocity evolution comparable to that of Si~{\sc ii}. The velocities measured from Si~{\sc ii} $\lambda5972$ and the S~{\sc ii} doublet are comparable to those of Si~{\sc ii} $\lambda 6355$, while the former two may show an increasing trend that is rarely seen in normal SNe~Ia after maximum light.

\section{Discussion}
\label{Section:discuss}
The optical light curves, color curves, and spectral evolution indicate that SN 2021hpr is a normal SN Ia, although its early light curves show excess emission at the beginning of the explosion. We will further explore the possible origins of the early excess emission in the context of a small sample with early observations. 

\subsection{The Distance and Luminosity}
\label{section:DL}
Different methods have been applied to estimate the distance to SN 2021hpr. 
We applied the {\texttt EBV model} of SNooPy2 to fit the light curves of SN
2021hpr in all bands and derived a distance modulus of $33.27 \pm 0.09$\,mag, where the uncertainty quoted is statistical. \cite{Zhang2022PASPhpr} and \cite{Lim2023} adopted the distance to SN 2021hpr as 33.46 $\pm$ 0.21 mag and 33.28 $\pm$ 0.11 mag, respectively.
Utilizing the observations of three siblings SNe Ia in NGC 3147 (including SN 1997pq, SN 2008qv, and SN 2021hpr), \cite{Ward2022} estimated the distance modulus as 33.14 $\pm$ 0.12 mag. \cite{Biscardi2012} calibrated the absolute peak magnitudes of SN 2008fv available in the literature for all bands, and found a distance modulus of 33.20 $\pm$ 0.10\,mag. Assuming an average distance modulus of the above estimates, $33.27 \pm 0.13$\,mag, the absolute \textit{B}-band peak magnitude of SN 2021hpr is $-19.16 \pm 0.14$\,mag, consistent with that of normal SNe~Ia \citep{Wang2009ApJ,Phillips1999AJ}.   

Following the methodology outlined by \citet{Li2019ApJ}, SNooPy2 is employed to establish the spectral energy distribution at various epochs and thus the quasibolometric light curve of SN 2021hpr based on the \textit{UBgVri} light curves. Around maximum light, both NIR and UV emissions are assumed to contribute $\sim 5$\% to the quasibolometric luminosity \citep{Wang2009ApJ,Zhang2016fe}. The peak luminosity of SN 2021hpr was then estimated as $\approx 1.11 \times 10^{43}\,\rm erg\,s^{-1}$ on MJD $59,322.35 \pm 0.58$ days, occurring $\sim  0.24$ days after the \textit{B}-band maximum. This peak luminosity is comparable to that of SN 2011fe ($1.13 \times 10^{43}\,\rm erg\,s^{-1}$; \citealp{Zhang2016fe}).

To estimate the ejecta parameters, we employ the Minim Code \citep{Chatzopoulos2013ApJ}, which is a modified radiation diffusion model of Arnett \citep{Arnett1982ApJ,Chatzopoulos2012ApJ,Li2019ApJ}. The Minim Code fits the quasibolometric light curves of SNe~Ia with a constant-opacity approximation. 
From the fit, the first light time ($t_0$) and the mass of radioactive $^{56}$Ni ejecta ($M_{\rm Ni}$) are 
estimated as MJD 59,305.35 $\pm$ 0.58 days and $0.57 \pm 0.05\,\rm M_{\odot}$, respectively. The model timescale of the light curve ($t_{\rm lc}$) is $15.71 \pm 0.01$ days, and the leaking timescale of gamma rays ($t_{\gamma}$) is $32.34 \pm 0.90$ days. According to the best-fitting results, the first light of Arnett's model is later than that estimated from the fireball model by $\sim 1.2$ days (see Section \ref{sec:FLT}). Note that the Arnett model does not take into account the "dark phase'' which is likely about 2 days  \citep{Piro2016ApJ}. Thus, the difference in the above two estimates is perhaps related to the dark phase \citep{Piro2016ApJ,Li2019ApJ,Zeng2021ApJ},or the above two models may not be suitable for measuring the FLT that have early excess samples. The mass of radioactive $^{56}$Ni synthesized in the explosion of SN 2021hpr is comparable to that of SN 2011fe ($M_{\rm Ni} = 0.53 \pm 0.11\,\rm M_{\odot}$; \citealt{Pereira2013}). Furthermore, using optimal $t_{\rm lc}$ and $t_{\gamma}$, we estimate the ejecta mass of SN 2021hpr as $0.83 \pm 0.05\,\rm M_{\odot}$ and the kinetic energy as $(0.75 \pm 0.09) \times 10^{51}$\,erg. These values are consistent with those of typical SNe~Ia \cite{Scalzo2019MNRAS}.


\subsection{Ratio of two Si~II Lines}
The depth ratio of Si~{\sc ii} $\lambda 5972$ and Si~{\sc ii} $\lambda 6355$, $R(\rm Si\,II)$, measured around the maximum light, has been proposed as an indicator of luminosity and/or temperature for SNe~Ia \citep{Nugent1995ApJ}. A lower value of $R(\rm Si\,II)$ generally corresponds to a more luminous SN~Ia with a higher photospheric temperature. For SN 2021hpr, the Si~{\sc ii} $\lambda 6355$ line was relatively broad at early times, while it became narrower around the time of maximum light. The Si~{\sc ii} $\lambda 5972$ line was visible in the spectra after $t \approx - 11.6$ days. 


Around the maximum light, the $R(\rm Si\,II)$ value of SN 2021hpr is small, indicating a high photospheric temperature. \citet{Benetti2005ApJ} noticed that $R(\rm Si\,II)$ shows a diverse evolution before maximum light, with the LVG SNe~Ia staying nearly constant and HVG SNe~Ia showing a downward trend, respectively. 
The intensity of Si~{\sc ii} $\lambda 4560$ is another temperature indicator of SNe~Ia, with stronger Si~{\sc ii} $\lambda 4560$ suggesting a higher photospheric temperature \citep{Benetti2004MNRAS}. The Si~{\sc ii} $\lambda 4560$ line in LVG SNe~Ia is found to be deeper than that of HVG SNe~Ia \citep{Pignata2008MNRAS}. As shown in Figures \ref{sepc1} and \ref{fig:sepc2}, the Si~{\sc ii} $\lambda 4560$ line in SN 2021hpr does not show significant evolution. And the shallow Si~{\sc ii} $\lambda 4560$ line perhaps suggests that the photospheric temperature of SN 2021hpr may not be so high, while this could be due to blending with Fe~{\sc ii}, Fe~{\sc iii}, and Mg~{\sc ii} emission lines \citep{Bongard2008ApJ,Yamanaka2009PASJ}. In this case, the mixed Si~{\sc ii} $\lambda 4560$ line cannot serve as a reliable indicator of the photospheric temperature of SNe~Ia.

\subsection{Origin of Early Excess Emission}
\label{Section:5.3}
\subsubsection{Companion shocking model}
During the phase immediately after the explosion, the multiband light-curve evolution can provide information on the progenitor system, the explosion mechanism, and even the circumstellar environment \citep{Kasen2010ApJ,Magee2020,Hu2023MNRAS,Li2023arXiv231114409L}. SN 2021hpr was discovered within $\sim 2.5$ days after the explosion, and our earliest observation began at $\sim 1.7$ days after the explosion. The color curves revealed the presence of relatively weak excess emission at early times (see Figure \ref{fig:color}), which may have a high-temperature component. The early quasi-bolometric light curve shows an excess emission of up to $\sim 7\%$ compared to the fireball model. To examine the origin of the observed flux excess, we utilize the \texttt{CompanionShocking3} model in the \texttt{lightcurve fitting} package to fit the early light curves \citep{Hosseinzadeh2022zndo}. This code employs the \texttt{emcee} package \citep{Foreman-Mackey2013PASPF}. 

\begin{figure} 
	\centering
	\includegraphics[height=3.in,width=\columnwidth]{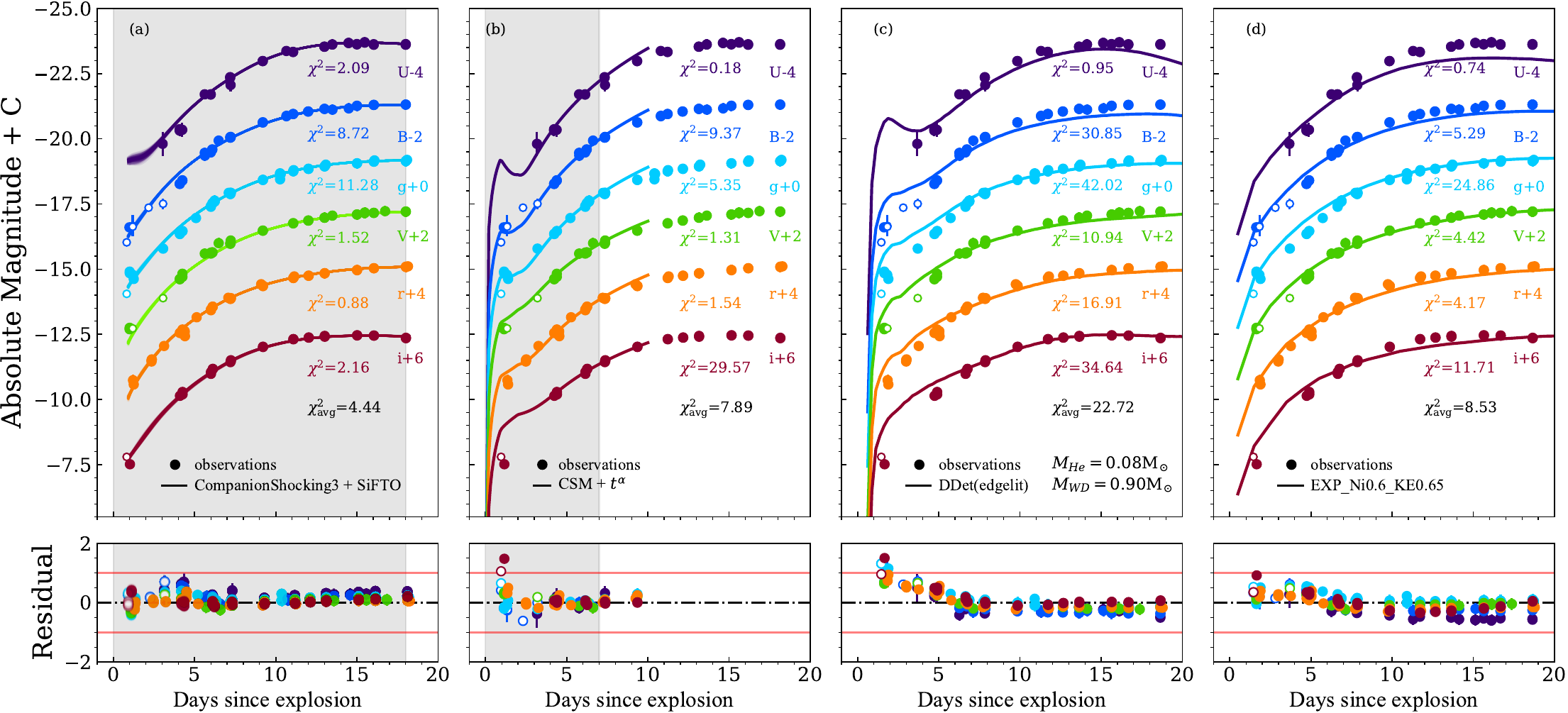}
	\caption{Fitting results of the early light curves (scatter) of SN 2021hpr using different models: (a) \texttt{CompanionShocking3} + \texttt{SiFTO} template model; (b) CSM + $t^{\alpha}$ model, with $\alpha$= 3.3, 2.9, 2.3, 2.7, 2.4, and 2.1 for the \textit{UBVgri} bands, respectively; (c) DDet model, the 0.08\,\rm $M_{\odot}$ mass of He shell detonation on the 0.9\,\rm $M_{\odot}$ mass WD model described by \cite{Polin2019ApJ}; and (d) the $^{56}$Ni mixing with the $\rm EXP\_Ni0.6\_KE0.65\_P4.4$ model described by \cite{Magee2020}. The residuals of the fitting curves are displayed in the lower panels. The unfilled points are the earliest data from \cite{Lim2023}. The chi-square of the residual is shown in the top panel. The chi-square values shown there were derived from the fitting results of the data from the first 7 days. The gray area of panels (a) and (b) represent the data involved in the empirical fitting}.
	\label{fig:lcc} 
\end{figure}

The \texttt{CompanionShocking3} model contains two components: \texttt{SiFTO} template with $s=1$ for SNe~Ia \citep{Conley2008ApJ}) and the companion shockinteraction component described by \cite{Kasen2010ApJ}. Because the \texttt{SiFTO} templates cover the \textit{UBVgri}-band data, we fit only the light curves of these bands. The fitting results include eight parameters: (1) the explosion time, $t_0$; (2) the binary separation of the companion-shocking component, $a$; (3) the viewing angle \citep{Brown2012ApJ}, $\theta$; (4) the time of the $B$-band peak for the \texttt{SiFTO} template, $t_{\rm max}$; (5) the stretch applied to the \texttt{SiFTO} component, $s$;  (6) a shift in the $U$-band maximum-light time for the \texttt{SiFTO} templates, $\Delta t_U$; (7) a shift in the $i$-band maximum time for the \texttt{SiFTO} templates, $\Delta t_i$.
The model parameters, along with their corresponding initial and best-fit values, are itemized in Table \ref{tab:kfit}. The light curves and the best-fit models are presented in Figure \ref{fig:lcc}a.
 
\startlongtable
\begin{deluxetable*}{lcccc}
\tablecolumns{6} 
\tablewidth{0pc} 
\tabletypesize{\scriptsize}
\tablecaption{\texttt{CompanionShocking3} model parameters}
\tablehead{\colhead{Parameter Variable$^{a}$} &\colhead{Units} &\colhead{Initial value$^b$} &\colhead{Best-fit value} }
\startdata
		  $t_0$             &MJD       &59,303.5, 59,305.0  & $59,304.10 \pm 0.02 $\\
            $a$               &$10^{13}\,\rm cm \approx 144\,\rm R_{\odot}$   &0, 1      & $0.11^{+0.03}_{-0.02}$ \\
  		$\theta$          &degree     &0, 180  &  $100\pm 10$\\
            $t_{\rm max}$     &MJD      &59,321.0, 59,323.0   & $59,321.41 \pm 0.02$\\
            $s$               &dimensionless         &0.5, 2   & $0.980 \pm 0.003$\\
            $\Delta t_U$      &days      &-1:1   & $0.34 \pm 0.02$\\
            $\Delta t_i$      &days      &-1:1   & $0.49 ^{+0.05}_{-0.04}$\\
\enddata
\tablenotetext{a}{See text for parameter descriptions.} 
\tablenotetext{b} {This column lists the maximum and minimum for a uniform distribution (separated by a comma) and a Gaussian distribution (separated by a colon).} 
\label{tab:kfit}
\end{deluxetable*}

Our analysis indicates that the \texttt{CompanionShocking3} model offers a better fit to the observation data. 
The optimal binary separation is determined as $15.84^{+4.32}_{-2.88}\,\rm R_{\odot}$, while the estimated radius of the companion star is $\sim 7.5\,\rm R_{\odot}$ (here we assumed Roche-lobe overflow in the fitting, \citealt{Eggleton1983ApJ}). This conclusion is consistent with the findings reported by \citet{Lim2023}. The best-fit explosion time is estimated as MJD $59,304.10 \pm 0.02$. In this paper, we adopted the average explosion time as MJD 59,304.13 $\pm$ 0.50 for SN 2021hpr, by taking into account the results from both the fireball model (Section \ref{sec:FLT}) and the \texttt{CompanionShocking3} model. Thus, the rise time of SN 2021hpr in \textit{B} is estimated as 17.98 $\pm$ 0.80 day. Since the Arnett model (Section \ref{section:DL}) does not consider a dark phase, we do not include it in the estimation of the explosion time. 

\subsubsection{Alternative explosion scenarios}
\begin{figure} 
	\centering
	\includegraphics[height=2.in,width=6.in]{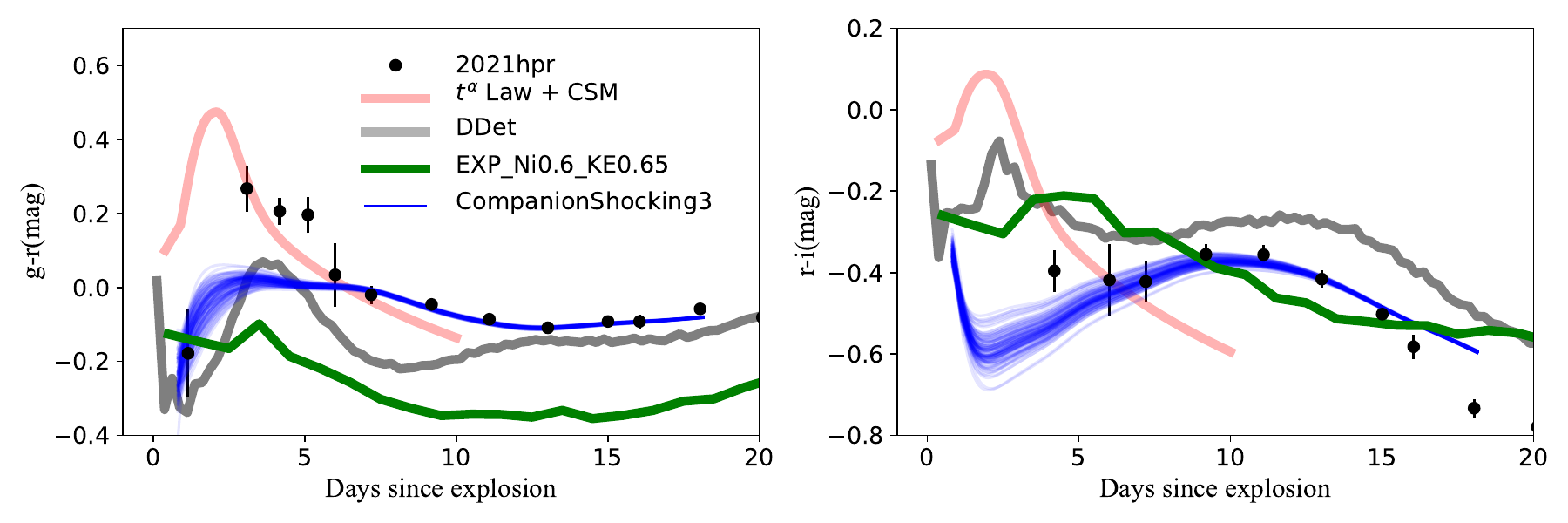}
	\caption{The $g-r$, and $r-i$ color curves of SN 2021hpr with different model color curves including \texttt{CompanionShocking3} + \texttt{SiFTO} template model; CSM + $t^{\alpha}$ model; DDet model, the $0.08\,\rm M_{\odot}$ mass of He shell detonation on the $0.9\,\rm M_{\odot}$ mass WD model described by \cite{Polin2019ApJ}; and the $^{56}$Ni mixing with $\rm EXP\_Ni0.6\_KE0.65\_P4.4$ model described by \cite{Magee2020}.}
	\label{fig:DD1} 
\end{figure}

The interaction between SN Ia ejecta and CSM may result in an excess of flux observed in early-time light curves. We employ the CSM model proposed by \citet{Hu2023MNRAS} to fit the early multiband light curves of SN 2021hpr and present the optimal fit results in Figure \ref{fig:lcc}b and \ref{fig:DD1}. This model assumes a CSM mass of $3.5 \times 10^{-5}\,\rm M_{\odot}$ around the SN at a distance of $5 \times 10^{14}\, \rm cm$, with a mass loss rate of $1.5 \times 10^{-6}\,\rm M_{\odot}\,yr^{-1}$. 
From the lower panel of Figure \ref{fig:lcc}b, one can see that the best-fit light curves exhibit relatively large deviations during the first two days, but they converge close to the observations in different bands at 4--10 days after the explosion. Thus, the CSM model cannot provide a satisfactory fit to the early-time excess emission seen in SN 2021hpr. 

An early flux excess in SNe~Ia can be attributed to the detonation of a thick He shell on a CO WD, due to radioactive materials present in the He-shell ashes, the so-called double-detonation (DDet) model \citep{Noebauer2017MNRAS.472.2787N,Jiang2017Natur.550...80J,Polin2019ApJ}. The DDet model posits that the ignition of the CO core is triggered by a thermonuclear explosion in the He shell, and the core explosion completely disrupts the entire WD \citep{Woosley1994ApJ}. In this case, the ashes of the He shell contain a significant amount of Fe-group elements that obstruct photons at shorter wavelengths and result in red colors at earlier phases. The early-time $r-i$ colors of SN 2021hpr are found to be redder than those of other normal types of SNe~Ia. On the other hand, a thick-shell DDet model is possibly consistent with such light curve and color-curve evolution.

\cite{Lim2023} compare the \textit{BVRI}-band light and color curves of SN 2021hpr with a thick He shell edge-lit DDet model ($0.9\,\rm M_{\odot}$ WD + $0.08\,\rm M_{\odot}$ He shell) of \citet{Polin2019ApJ}, and conclude that DDet models do not provide a perfect explanation of its early-time light curves. We used the same model to conduct a comparative analysis, as shown in Figure \ref{fig:lcc}c and Figure \ref{fig:DD1}. The color-curve shapes given by the model are similar to the observations, the $g-r$ color is bluer than the observation in the first 4 days and the $r-i$ color is redder in the first 20 days. After that, the DDet model produces a color curve that is too red. Thus, we reach the same conclusion as \cite{Lim2023} that the predictions of DDet are not fully consistent with the observed properties of SN 2021hpr, as also indicated by the large $\chi^{2}$ in the fit (see Fig. 10c).  

The distribution of $^{56}$Ni within the ejecta of SNe~Ia can significantly affect the early-epoch light curves. We compared the model set calculated by \cite{Magee2020} and \cite{Magee2020A&A} with the early light curves of SN 2021hpr. Among the models in the set, the $\rm EXP\_Ni0.6\_KE0.65\_P4.4$ model shows the highest degree of similarity to the observed early light curves of SN 2021hpr, as shown in Figure \ref{fig:lcc}d. In this model, the distributions of $^{56}$Ni are shown for an exponential density profile, a kinetic energy of $6.53 \times 10^{50}\, \rm erg$, with a $^{56}$Ni mass of $0.6\,\rm M_{\odot}$ \citep{Magee2020A&A}. It should be noted that the $^{56}$Ni mass of SN 2021hpr calculated by the Arnett model is $0.57 \pm 0.05\,\rm M_{\odot}$ (see Section \ref{section:DL}), which is consistent with the mass given in the $\rm EXP\_Ni0.6\_KE0.65\_P4.4$ model. 
However, the model color curves are generally inconsistent with the observations, except for the $r-i$ color at late times, as shown in Figure \ref{fig:DD1} (see green lines). 

It should be noted that the companion and CSM interaction models are empirical fits to the data, while the DDet and $^{56}$Ni mixing models are obtained from model grids.
From the average Chi-square($\chi^2$) of the residuals, ‌it can be found that the \texttt{CompanionShocking3} model provides a better fit‌ to the early light curves of SN 2021hpr compared to the other three models. However, the viewing angle provided by this model's fit only marginally satisfies the identification criteria proposed by \cite{Burke2022} within the error range. The observed $g-r$ and $r-i$ color curves (Figure \ref{fig:DD1}) do not exhibit a perfect match with this model. Therefore, it cannot be ruled out that other possible models could also explain the flux excess observed in the early stages of SN 2021hpr.

\subsubsection{Spectroscopic differences in SNe Ia with and without early-excess emission}
\begin{figure} [ht!]
	\centering
	\includegraphics[height=5.5in,width=4.in]{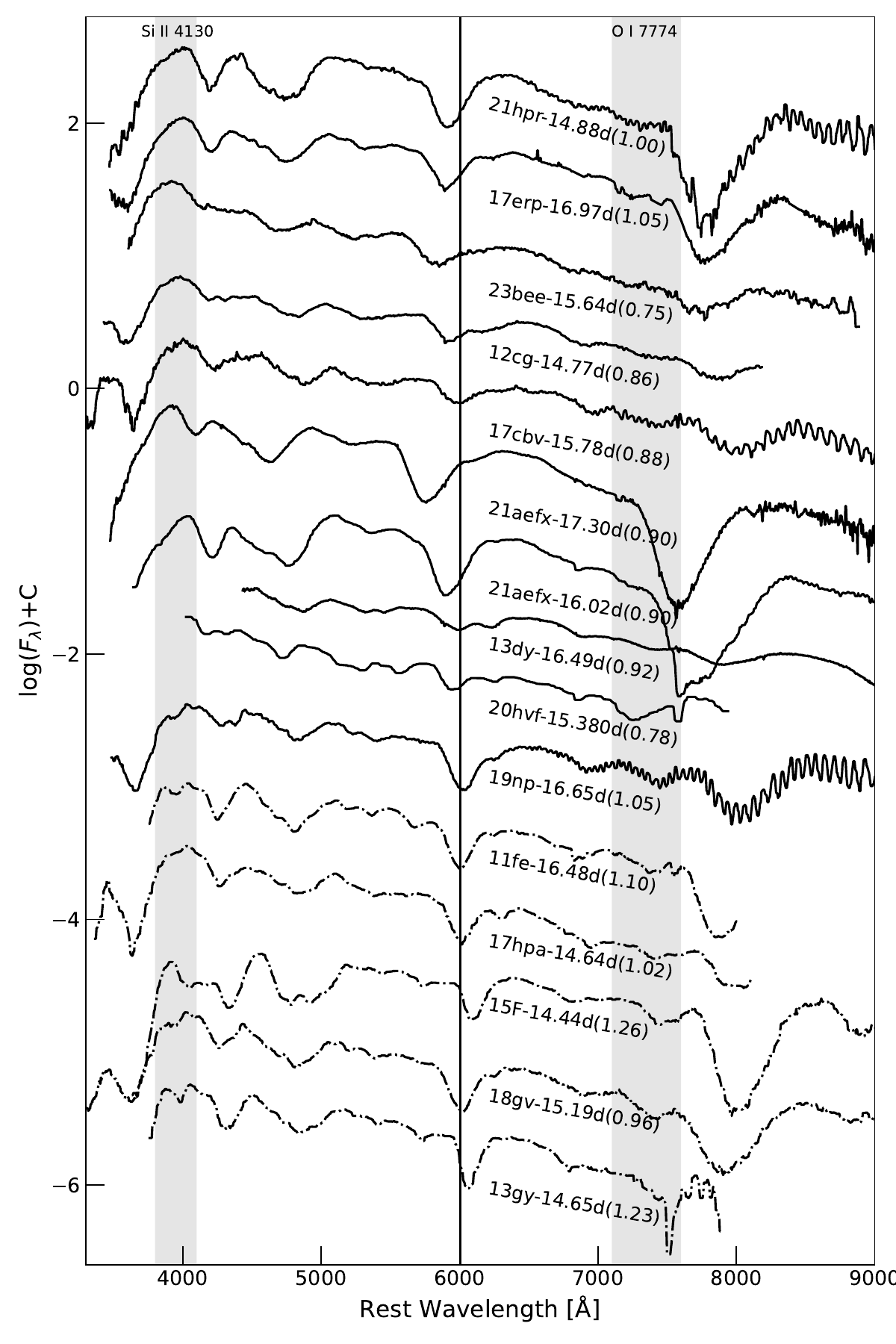}
	\caption{Comparison of the earliest spectra of SN 2021hpr, SN 2017erp \citep{Burke2022}, SN 2019np \citep{Sai2022MNRAS}, SN 2017cbv \citep{Wee2018ApJ} ,SN 2023bee \citep{Hosseinzadeh2023ApJ}, SN 2012cg \citep{Marion2016ApJ}, SN 2021aefx \citep{Hosseinzadeh2022ApJ,Ni2023arXiv}, SN 2015F \citep{Cartier2017MNRAS}, SN 2018gv \citep{Yang2020ApJ}, SN 2020hvf \citep{Jiang2021ApJ}, SN 2011fe \citep{Zhang2016fe}, SN 2017hpa \citep{Zeng2021ApJhpa}, SN 2013gy \citep{Holmbo2019A&A}, and SN 2013dy \citep{Zheng2013ApJ}. The solid lines represent SNe~Ia with early flux excess, and the dashed lines represent those without early flux excess. The vertical line corresponds to the absorption minimum of the Si~{\sc ii} $\lambda 6355$ absorption line of SN 2011fe. The label for each line includes the abbreviated name of the SN~Ia, its phase relative to \textit{B}-band maximum light, and (in parentheses) its $\Delta m_{15}(B)$ value. The gray area is marked to indicate the spectral features at around the  Si~{\sc ii} $\lambda 4130$ and O~{\sc i} $\lambda 7774$ absorption lines.}
	\label{fig:espec} 
\end{figure}

Regardless of which theoretical model (e.g., \texttt{CompanionShocking} model; CSM + $t^{\alpha}$ model, DDet model;  $^{56}$Ni mixing model, etc.) is responsible for the early flux excess, the early light curves and spectra will exhibit some qualitative clues about the progenitor stars and/or the explosion mechanism of SNe~Ia. In Figure \ref{fig:espec}, we compare the earliest spectra of SNe~Ia with (YEs) and without (NEs) early flux excess. We selected the normal type SNe Ia that have been firmly identified for the presence or absence of early excess emission in literature. Moreover, early spectra of at least two weeks before maximum light should be available for this sample
including SN 2017erp \citep{Burke2022}, SN 2019np \citep{Sai2022MNRAS}, SN 2017cbv \citep{Wee2018ApJ} ,SN 2023bee \citep{Hosseinzadeh2023ApJ}, SN 2012cg \citep{Marion2016ApJ}, SN 2021aefx \citep{Hosseinzadeh2022ApJ,Ni2023arXiv}, SN 2020hvf \citep{Jiang2021ApJ},  SN 2013dy \citep{Zheng2013ApJ}, SN 2015F \citep{Cartier2017MNRAS}, SN 2018gv \citep{Yang2020ApJ}, SN 2011fe \citep{Zhang2016fe}, SN 2017hpa \citep{Zeng2021ApJhpa}, and SN 2013gy \citep{Holmbo2019A&A}.

\begin{figure}[ht!] 
	\centering
	\includegraphics[height=3.in,width=6in]{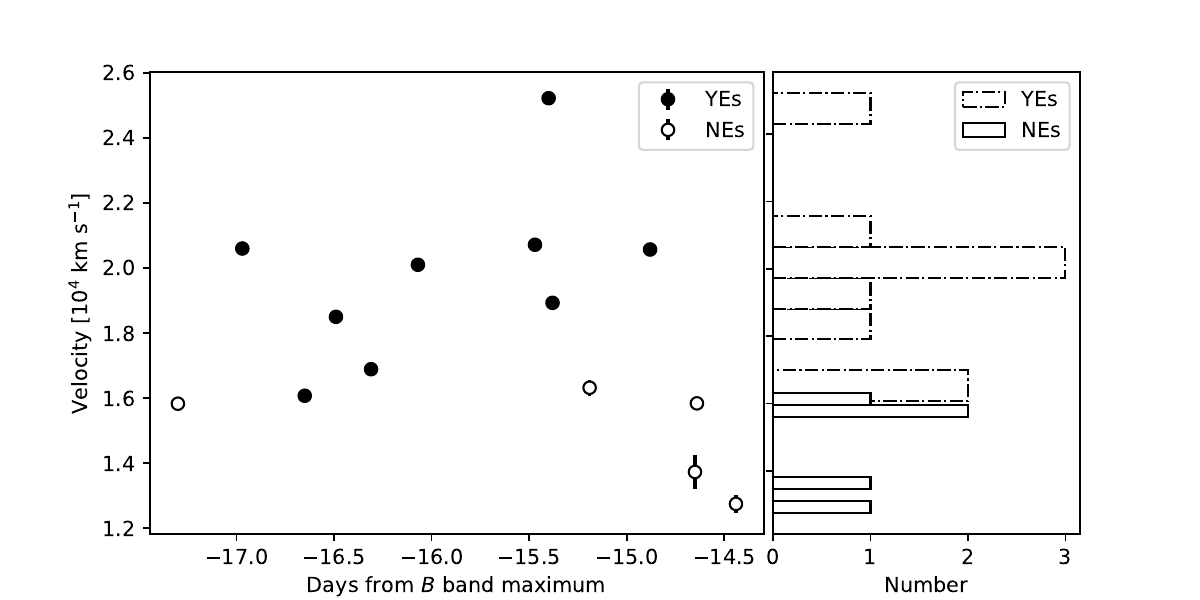}
	\caption{Left panel: early expansion velocity distribution of 14 normal SNe Ia. The observed sample is the same as Figure \ref{fig:espec}, but solid circles and open circles represent the YEs samples and the NEs, respectively. Right panel: the corresponding histogram for the early expansion velocity}
	\label{fig:evelocity} 
\end{figure}
We found significant differences in the earliest spectra of YEs and NEs SNe~Ia. The NEs SNe~Ia show prominent Si~{\sc ii} $\lambda 4130$ and O~{\sc i} $\lambda 7774$ absorption lines, but those YEs SNe~Ia do not or exhibit substantially weaker lines (except for SN 2019np). The \texttt{CompanionShocking} model is hardly able to explain the flux excess of SN 2019np, but the $^{56}$Ni mixing model is more consistent with this SN \citep{Sai2022MNRAS,Burke2022}. In Figure \ref{fig:espec}, the vertical line corresponds to the absorption minimum of the Si~{\sc ii} $\lambda 6355$ line of SN 2011fe, corresponding to a velocity of $\sim 15,900\,\rm km\, s^{-1}$ \citep{Zhang2016fe}. The absorption minimum of Si~{\sc ii} $\lambda 6355$ in YE SNe~Ia is bluer than that of the NEs SNe~Ia, indicating that the YE objects may have higher ejecta velocities than the NE objects at early phase. Figure \ref{fig:evelocity} displays the distribution of the early-phase Si II velocity measured for a sample of normal SNe Ia with early observations. To investigate whether the YEs and NE samples come from different groups, based on their early-time velocities, we performed a K-S test and obtained the $P$ value as $6 \times 10^{-3}$. This low value suggests a significant difference between the velocities of YEs and NEs objects in the very early phase. 
Discrepancy can be also seen in their photometric parameters. We found that the average decline rate of the YEs sample is noticeably smaller than that of the NEs sample (i.e., 0.91$\pm$0.01 versus $1.11\pm0.03$ mag). This is not unexpected since the YEs SNe Ia should have additional energy sources, perhaps due to interactions with CSM/companion stars or more complete burning at outer layers.
However, it should be noted that the above results could be affected by the limited sample available in the literature, and a more extensive dataset is required for a more thorough analysis. Most of these YEs SNe~Ia can be put in the NV subclass according to their Si II velocity measured at around the maximum light, except for SN 2023bee which has a velocity $12,150 \pm 50\,\rm km\,s ^{-1}$ \citep{Hosseinzadeh2023ApJ} at this phase. This indicates that the YEs SNe~Ia tend to have detached HVFs at an early stage, but evolve like NV SNe Ia when approaching maximum light.  


\subsection{Late-Time Spectra}
\begin{figure}[ht!] 
	\centering
	\includegraphics[height=4.in,width=5in]{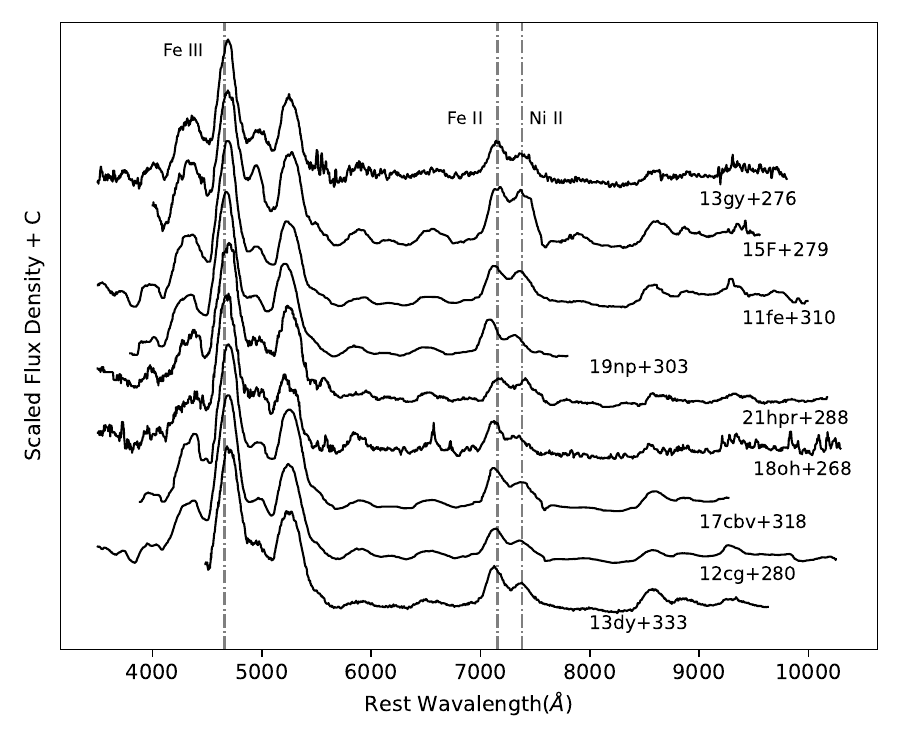}
	\caption{Late-time spectra of SNe 2011fe, 2012cg, 2013dy, 2013gy, 2015F, 2017cbv, 2018oh, 2019np, and 2021hpr. Flux densities are normalized to the [Fe~{\sc iii}]\,$\lambda 4701$, and smoothed with a bin of 25 $\mathring{\rm A}$. The three vertical dash-dotted lines from left to right represent the rest-frame wavelengths of [Fe~{\sc iii}]\,$\lambda 4701$, [Fe~{\sc ii}]\,$\lambda 7155$, and [Ni~{\sc ii}]\,$\lambda 7378$, respectively.}
	\label{fig:hprneb} 
\end{figure}

Nebular spectra of SNe~Ia can provide powerful probes of the underlying physics of the explosion \citep{Liu2023MNRAS,Graham2022MNRAS,Maguire2018MNRAS,Black2016MNRAS}. Redshifted or blueshifted nebular velocities of SNe~Ia might have a connection with the explosion geometry \citep{Maeda2010ApJ}. The nebular velocity represents the average velocity of [Fe~{\sc ii}]\,$\lambda 7155$ and [Ni~{\sc ii}]\,$\lambda 7378$ velocities \citep{Maeda2010ApJ,Silverman2013ApJS}. 
Figure \ref{fig:hprneb} presents late-time spectra of SN 2021hpr alongside those of other well-studied normal SNe Ia,  including SNe Ia 2011fe, 2012cg, 2013dy, 2013gy, 2015F, 2017cbv, 2018oh, 2019np, and 2021hpr. The shape of the late-time spectra of SN 2021hpr is extremely similar to that of other comparison SNe Ia. However, the main difference lies in the intensity evolution of individual emission lines. In particular, the emission line intensity of [Fe~{\sc iii}]\,$\lambda 4701$ tends to decrease over time, as observed in cases such as SN 2011fe and SN 2017cbv; while in the case of SN 2021hpr, this emission line tends to become stronger with time. Furthermore, in the even later spectrum of SN 2021hpr, the intensity of [Fe~ {\sc ii}]\,$\lambda 7155$ and [Ni~{\sc ii}]\,$\lambda 7378$ lines became comparable. As the ejecta expands, it becomes more transparent, and the radiation caused by the decay of the innermost iron-group elements appears to be stronger.

\begin{figure} 
	\centering
	\includegraphics[height=3.in,width=3.in]{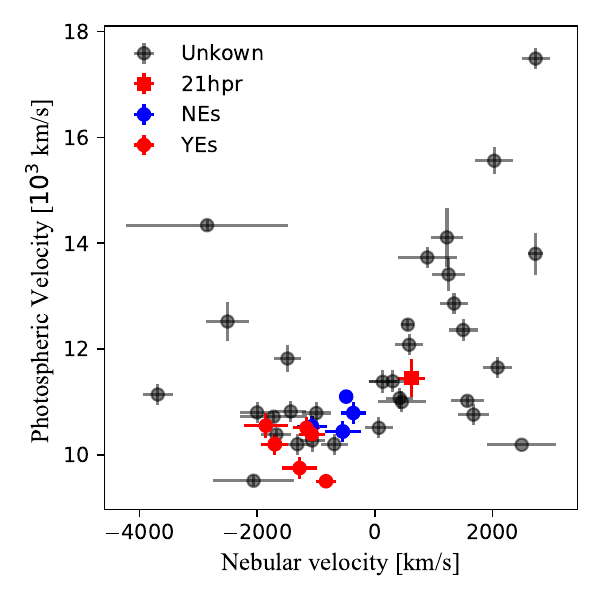}
	\caption{The photospheric velocity measured at maximum light versus the velocity shift inferred from the nebular-phase spectra. The red square represents SN 2021hpr measured using Multi-Gaussian fits to the Fe~{\sc ii} and Ni~{\sc ii} blended features at $t\approx 288$ days. Data from \cite{Maeda2010ApJ,Silverman2012MNRAS,Graham2022MNRAS,Liu2023MNRAS}.}
	\label{fig:vmaxvsvneb} 
\end{figure}

To obtain the velocities of the emission lines in these spectra, we used the direct measurement method for the forbidden emission lines of [Fe~{\sc iii}]\,$\lambda 4701$ and [Fe~{\sc ii}]\,$\lambda 5250$. In contrast, the [Fe~ {\sc ii}]\,$\lambda 7155$ and [Ni~{\sc ii}]\,$\lambda 7378$ lines are fitted with the multi-Gaussian method. The measured velocities are presented in Table \ref{nebularvelocity}. 
The estimated velocities of [Fe~{\sc ii}] $\lambda 5250$, [Fe~{\sc ii}] $\lambda 7155$, and [Ni~{\sc ii}] $\lambda 7378$ show a redshift evolution, while  [Fe~{\sc iii}] $\lambda 4701$ exhibits a blueshift trend.
The [Ni~{\sc ii}]\,$\lambda 7378$ line is weaker than [Fe~{\sc ii}]\,$\lambda 7155$ at $t\approx +263$ days. Thus, we use [Fe~{\sc ii}]\,$\lambda 7155$ to represent the nebular velocity at this phase. For SN 2021hpr, this velocity shift is calculated as 710 $\pm$ 170\,$\rm km\, s^{-1}$ at $t\approx 263$ days and 640 $\pm$ 100\,$\rm km \, s^{-1}$ at 
$t\approx +288$ days, respectively, suggesting a redshifted nebular velocity for this SN. Figure \ref{fig:vmaxvsvneb} shows the photospheric velocity measured around the maximum light versus the velocity shift measured from Fe~{\sc ii} and Ni~{\sc ii} lines in the nebular phase. The velocities of the [Fe~{\sc iii}]\,$\lambda 4701$, [Fe~{\sc ii}]\,$\lambda 5250$, and [Ni~{\sc ii}]\,$\lambda 7378$ emission lines exhibit a redward evolution over time, consistent with that found for normal SNe~Ia \citep{Black2016MNRAS,Maguire2018MNRAS,Graham2022MNRAS}. 

\begin{figure} 
	\centering
	\includegraphics[height=3.in,width=5.in]{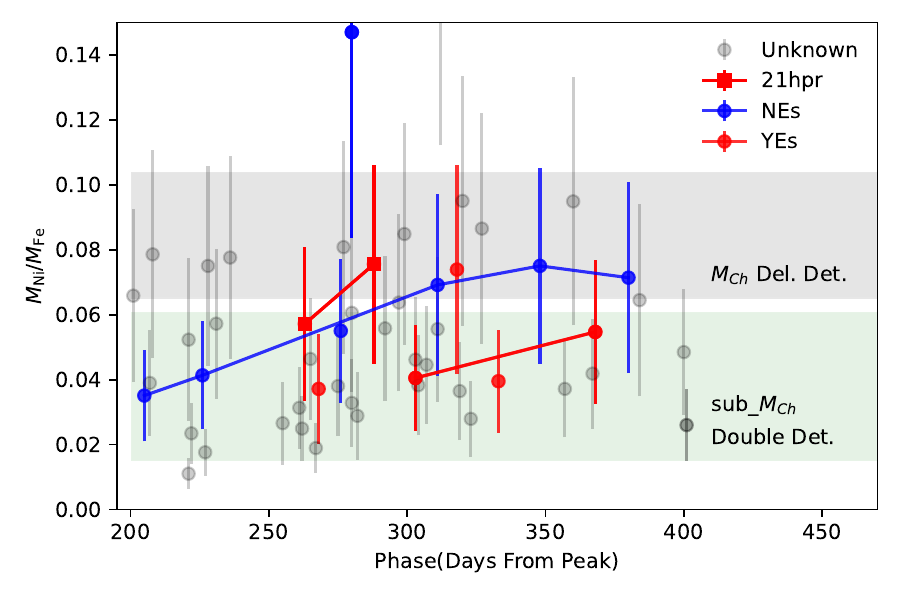}
	\caption{Mass ratio distribution of Ni and Fe estimated by the late-time spectra of normal SNe Ia. Data from  \cite{Liu2023MNRAS}. The gray area for DDT models \citep{Seitenzahl2013MNRAS}, and the green area for sub-$M_{\rm Ch}$ models \citep{Shen2018ApJ}.}
	\label{fig:nivsfe} 
\end{figure}

From $t\approx 263$ days to $t\approx 288$ days, the [Ni~{\sc ii}]\,$\lambda 7378$ line becomes relatively more prominent, leading to an increased Ni/Fe ratio. This change is likely due to that the inner ejecta cool gradually over time, as seen in other SNe~Ia \citep{Liu2023MNRAS,Blondin2022A&A}. The Ni/Fe ratio, the flux ratio of [Fe~{\sc ii}] $\lambda 7155$ and [Ni~{\sc ii}] $\lambda 7378$, is used to constrain the explosion mechanism of SNe Ia \citep{Maguire2018MNRAS}.
Figure \ref{fig:nivsfe} shows the result for SN 2021hpr, however, this change in Ni/Fe ratio due to the spectral evolution makes it move from the sub$-M_{\rm Ch}$ double-detonation model region to the Chandrasekhar $M_{\rm Ch}$ delayed-detonation-model region. This indicates that constraining the explosion model based on the Ni/Fe ratio is still challenging for an individual SN Ia. ‌

We tried to examine distributions of the mass ratio of Ni to Fe estimated from the late-time spectra of normal SNe Ia (see Figure \ref{fig:nivsfe}), including YEs and NEs objects, but found no signficant tendency between these two subgroups. More samples of SNe Ia with both very early and nebular phase observations are needed for better quantitative analysis. 

\begin{table}
	\centering
	\setlength{\tabcolsep}{4mm}
	\caption{Velocities of nebular-phase emission lines in SN 2021hpr}
	\begin{tabular}{llll}
		\hline
		UT Date & Phase & Line & Velocity\\
		       & [days]&      & [$\rm km\, s ^{-1}$]\\
		\hline
		2022-01-06 & 263 & [Fe {\sc iii}]\,$\lambda 4701$ & -1100$\pm$60\\
		2022-01-06 & 263 & [Fe {\sc ii}]\,$\lambda 5250$  & 1200$\pm$100\\
		2022-01-06 & 263 & [Fe {\sc ii}]\,$\lambda 7155$  & 710$\pm$170\\
		2022-01-06 & 263 & [Ni {\sc ii}]\,$\lambda 7378$  & 120$\pm$120\\
		\hline
		2022-01-30 & 288 & [Fe {\sc iii}]\,$\lambda 4701$ & -1050$\pm$30\\
		2022-01-30 & 288 & [Fe {\sc ii}]\,$\lambda 5250$  & 1330$\pm$50\\
		2022-01-30 & 288 & [Fe {\sc ii}]\,$\lambda 7155$  & 540$\pm$150\\
		2022-01-30 & 288 & [Ni {\sc ii}]\,$\lambda 7378$  & 730$\pm$130\\
		\hline
	\end{tabular}	
    \label{nebularvelocity}
\end{table}


\section{Conclusion}
\label{section:conclusion}
We present comprehensive photometric and spectroscopic observations of SN 2021hpr, covering the phase from about 1 day to 290 days after the explosion. The main photometric and spectroscopic parameters derived for SN 2021hpr are listed in Table \ref{mianpar}.
Spectroscopically, it is a normal type Ia supernova, while its early-time light curves reveal faint excess emission.
Based on the expansion velocity near the maximum light and post-peak velocity gradient, SN 2021hpr can be categorized into the NV and LVG subclasses. The ejecta velocity of SN 2021hpr was found to undergo a drastic decline when approaching the maximum light, with a gradient of 571$\pm$18\, $\rm km\, s^{-1}\,day^{-1}$, which is larger than normal SNe Ia like SN 2011fe.

Among different models proposed to account for the early excess emission in SNe Ia, the \texttt{CompanionShocking3} model provides better fit to the early data. This model gives the binary separation as $\sim 15.84\,\rm R_{\odot}$ and a companion radius of $\sim 7.5\,\rm R_{\odot}$. These results are consistent with previous studies. The fitting results of the companion interaction model favor the existence of a nondegenerate companion in the progenitor system of SN 2021hpr. Alternatively, we also employ the DDet, CSM, and $^{56}$Ni mixing models to discuss the early flux excess of SN 2021hpr. It is difficult for both the DDet and $^{56}$Ni mixing models to reproduce the early-phase light and color curves of SN 2021hpr. 
 
With a small sample of well-observed normal SNe Ia, we find that the appearance of early excess emission in the light curves is likely related to a large velocity gradient inferred from the spectra. This favors a physical origin of the  interaction of ejecta with non-degenerate companion and/or surrounding CSM for SNe Ia like SN 2021hpr. For SN 2021hpr, however, the Fe~{\sc ii}/Ni~{\sc ii} lines are found to show a redshifted velocity as inferred from its nebular-phase spectra, while this velocity is blueshifted for all the comparison SNe Ia showing early bump features. More statistical sample with better observations in both early and nebular phases are needed to clarify whether SN 2021hpr is an outerlier in this respect.   
\begin{table}[ht!]
	\centering
	\setlength{\tabcolsep}{2mm}
	\caption{Comparison of main parameters of SN 2021hpr and SN 2011fe}
	\begin{tabular}{lllll}
		\hline
		Parameter & Unit &SN 2021hpr$^A$ & SN 2011fe &SN 2021hpr $^P$\\
		\hline
		$\Delta m_{15}(B)$         & mag     &1.00 $\pm$ 0.01        &1.18 $\pm$ 0.03$^{Z}$    & 0.988 $\pm$ 0.026$^{L}$\\
		$t_{\rm max}(B)$           & days    &59,322.11 $\pm$ 0.58   &55,815.5 $\pm$ 0.30      &59,321.856 $\pm$ 0.218$^{L}$\\
		$B_{\rm max}$              & mag     &14.11 $\pm$ 0.04       &10.0 $\pm$ 0.02          &14.017 $\pm$ 0.017$^{Z2}$ \\
		$B_{\rm max}-V_{\rm max}$  & mag     &-0.028 $\pm$ 0.007     &-0.03 $\pm$ 0.04$^{Z}$   &-0.004 $\pm$ 0.005$^{L}$\\
		$M_{\rm max}(B)$           & mag     &-19.16 $\pm$ 0.14      &-19.21 $\pm$ 0.15        &-19.553 $\pm$ 0.111$^{L}$\\
		$E(B-V)_{\rm host}$        & mag     &0.06 $\pm$ 0.06       & 0.032 $\pm$ 0.045$^{Z}$ &0.079 $\pm$ 0.040$^{L}$\\
		$s_{BV}$                   & dimensionless        &1.02 $\pm$ 0.03          & -        &None\\
		$t_0$                      & days    &59,304.13 $\pm$ 0.50   &55,796.48$\pm$0.16$^{Z,Z1}$  &59,304.73 $\pm$ 0.01$^{L}$ \\
            $\rm DM$               & mag     &33.27 $\pm$ 0.13       &29.04 $\pm$ 0.05$^{S}$       &33.28 $\pm$ 0.11$^{L}$ \\
		$\tau_{\rm rise}$          & days    &17.98 $\pm$ 0.80       &18.00 $\pm$ 0.16$^{Z,Z1}$    &16.424 $\pm$ 0.078$^{Z3}$ \\
		$L_{\rm bol}^{\rm max}$    & $\rm erg\, s^{-1}$   &$\approx 1.11 \times 10^{43}$    &(1.13$\pm$0.07)$\times 10^{43}$$^{Z}$ &None\\
		$M_{\rm ^{56}Ni}$          & $\rm M_{\odot}$        &0.57 $\pm$  0.05  &0.53 $\pm 0.11^{S}$  &0.44 $\pm$  0.14$^B$\\
		\hline
		$\upsilon_{\rm 0}$(Si {\sc ii}) & km s$^{-1}$     &11,453   $\pm$ 100 &10,400$^{Z}$         & $\sim$ 12,420$^{Z2,Z4}$ \\
		$R$(Si {\sc ii})                &dimensionless    &0.08 $\pm$ 0.01 & 0.18$\pm$ 0.02$^{Z5}$  &None\\
		$\dot{\upsilon}$(Si {\sc ii})   & km s$^{-1} \rm \, day^{-1}$        &18 $\pm$ 6     & 52.4$^{Z}$ &None\\
		\hline
	\end{tabular}
 
\raggedright $^{Z}$ \cite{Zhang2016fe}, $^A${This work}, $^P${Previous} , $^{Z1}${for the $t^n$ model}, $^{S}$\cite{Shappee2011ApJ}, $^{L}$\cite{Lim2023}, $^{Z2}$\cite{Zhang2022PASPhpr}, $^{Z3}${for the \textit{B} band, and $t_0^B$ = 59,305.438 $\pm$ 0.450, $t_{\rm max}^B$ = 59,321.862 $\pm$ 0.450}, $^{Z4}${four days before maximum brightness}, $^{Z5}$ \cite{Pereira2013}, $^{B}$ \cite{Barna2023AA}.
\label{mianpar} 
\end{table}

\section*{Acknowledgements}
This work is sponsored by the National Natural Science Foundation of China (NSFC grants 12288102, 12033003, 12203029, 12373038, 11803076, and 12433007), Natural Science Foundation of Xinjiang Uygur Autonomous Region under No. 2024D01D32, Tianshan Talent Training Program (grant 2023TSYCLJ0053, 2023TSYCCX0101), the Central Guidance for Local Science and Technology Development Fund under No. ZYYD2025QY27, the New Cornerstone Science Foundation through the XPLORER PRIZE, the Chinese Academy of Sciences (CAS) "Light of West China" Program (grant 2020-XBQNXZ-016), the Strategic Priority Research Program of the Chinese Academy of Sciences (grants XDB0550100, XDB0550000), and the High-Level Talent-Heaven Lake Program of Xinjiang Uygur Autonomous Region of China, the NKFIH/OTKA FK-134432 grant of the National Research, Development and Innovation (NRDI) Office of Hungary. This work includes data from the Las Cumbres Observatory global telescope network; the LCO group is supported by NSF grants AST-1911151 and AST-1911225.

L.G. acknowledges financial support from AGAUR, CSIC, MCIN and AEI 10.13039/501100011033 under projects PID2023-151307NB-I00, PIE 20215AT016, CEX2020-001058-M, ILINK23001, COOPB2304, and 2021-SGR-01270.
J.Z. is supported by the National Natural Science Foundation of China (grant 12173082), the Yunnan Province Foundation (grant 202201AT070069), the Top-notch Young Talents Program of Yunnan Province, the Light of West China Program provided by the Chinese Academy of Sciences, the International Centre of Supernovae, Yunnan Key Laboratory (grant 202302AN360001).
A.V.F.’s research group at UC Berkeley acknowledges financial assistance from the Christopher R. Redlich
Fund, Gary and Cynthia Bengier, Clark and Sharon Winslow, Alan Eustace (W.Z. is a Bengier-Winslow-Eustace Specialist in Astronomy), William Draper, Timothy and Melissa Draper, Briggs and Kathleen Wood, Sanford Robertson (T.G.B. is a Draper-Wood-Robertson Specialist in Astronomy), and numerous other donors.                  
A major upgrade of the Kast spectrograph on the Shane 3\,m telescope at Lick Observatory, led by Brad Holden, was made possible through generous gifts from the Heising-Simons Foundation, William and Marina Kast, and the University of California Observatories. Research at Lick Observatory is partially supported by a generous gift from Google.  
For their expert assistance, we acknowledge the staff of the LCO telescopes, the Nanshan One-meter Wide-field Telescope, the Lijiang 2.4\,m telescope, the Lick Observatory Shane 3\,m telescope, the twin Keck 10\,m telescopes, the Asteroid Terrestrial-impact Last Alert System (ATLAS) project, and the BFOSC mounted on the Xinglong 2.16\,m telescope. Some of the data presented herein were obtained at the W. M. Keck
Observatory, which is operated as a scientific partnership among the
California Institute of Technology, the University of California, and
NASA; the observatory was made possible by the generous financial
support of the W. M. Keck Foundation.

\appendix

\setcounter{table}{0}
\renewcommand{\thetable}{A\arabic{table}}
We present photometric datasets of SN 2021hpr in three tables: Table \ref{tab:atlas_o} lists measurements obtained with LCO, NOWT, ZTF, and ATLAS; Table \ref{tab:kast} contains data from the KAIT; and Table \ref{tab:lc_clear} documents unfiltered optical photometry acquired through the 0.35 m telescope at Itagaki Astronomical Observatory.

\startlongtable
\begin{deluxetable*}{lcccll}
\tablecolumns{3} 
\tablewidth{5pc} 
\tabletypesize{\scriptsize}
\tablecaption{Photometry of SN 2021hpr taken with LCO, NOWT, ZTF, and ATLAS}
\tablehead{\colhead{MJD} &\colhead{Epoch$^a$} &\colhead{Mag} &\colhead{Magerr}&\colhead{Telescope}&\colhead{Filter}}
\startdata
59,305.089 	&-17.021 	&18.776 	&0.152 	&LCO	     &B\\
59,305.099 	&-17.011 	&18.661 	&0.130 	&LCO	     &V\\
59,305.102 	&-17.009 	&18.608 	&0.126 	&LCO	     &V\\
59,305.104 	&-17.006 	&18.468 	&0.090 	&LCO	     &g\\
59,305.108 	&-17.002 	&18.526 	&0.092 	&LCO	     &g\\
59,305.120 	&-16.990 	&19.814 	&0.079 	&LCO	     &i\\
59,305.285 	&-16.825 	&18.591 	&0.235 	&ZTF	     &g\\
59,305.309 	&-16.801 	&18.607 	&0.229 	&ZTF	     &r\\
59,305.353 	&-16.758 	&18.748 	&0.104 	&LCO	     &g\\
59,305.354 	&-16.756 	&18.769 	&0.119 	&LCO	     &r\\
..&..&..&..&..&..\\
59,531.388 	&209.278 	&19.269 	&0.119 	&ZTF	     &g\\
59,531.408 	&209.298 	&20.452 	&0.292 	&ZTF	     &r\\
59,538.366 	&216.256 	&19.272 	&0.238 	&ZTF	     &g\\
59,550.429 	&228.319 	&19.549 	&0.095 	&ZTF	     &g\\
59,588.550 	&266.440 	&19.734 	&0.228 	&ATLAS	     &c\\
\enddata
\tablenotetext{a}{Relative to the epoch of $B$-band maximum brightness (MJD = 59,322.11).}
\tablenotetext{}{This table is available in its entirety in machine-readable form.}
\label{tab:atlas_o}
\end{deluxetable*}

\startlongtable
\begin{deluxetable*}{lccccccccccc}
\tablecolumns{6} 
\tablewidth{0pc} 
\tabletypesize{\scriptsize}
\tablecaption{Photometry of SN 2021hpr taken with KAIT}
\tablehead{\colhead{MJD} &\colhead{Mag $B$} &\colhead{Magerr $B$}&\colhead{Mag $V$} &\colhead{Magerr $V$}&\colhead{Mag $R$} &\colhead{Magerr $R$}
&\colhead{Mag $Clear$} &\colhead{Magerr $Clear$}&\colhead{Mag $I$} &\colhead{Magerr $I$}}
\startdata
59,309.241	&16.783	 &0.054	&16.269	 &0.032	&16.229	 &0.068	&16.144	 &0.049	&16.368	 &0.096\\
59,310.294	&16.031	 &0.077	&15.794	 &0.103	&15.776	 &0.210	&15.677	 &0.055	&15.912	 &0.194\\
59,312.206	&15.386	 &0.045	&15.238	 &0.038	&15.222	 &0.079	&15.069	 &0.062	&15.345	 &0.047\\
59,313.298	&15.165	 &0.047	&15.004	 &0.042	&14.938	 &0.050	&14.816	 &0.039	&15.064	 &0.090\\
59,314.208	&15.014	 &0.146	&14.874	 &0.176	&14.820	 &0.279	&14.613	 &0.292	&14.778	 &0.852\\
59,315.261	&14.656	 &0.059	&14.676	 &0.028	&14.633	 &0.026	&14.525	 &0.046	&14.735	 &0.036\\
59,316.266	&14.705	 &0.035	&14.599	 &0.020	&14.550	 &0.024	&14.423	 &0.038	&14.693	 &0.024\\
59,317.329	&14.523	 &0.058	&14.458	 &0.036	&14.441	 &0.047	&14.348	 &0.037	&14.631	 &0.097\\
59,318.338	&14.405	 &0.095	&14.410	 &0.033	&14.375	 &0.032	&14.257	 &0.037	&14.599	 &0.039\\
59,320.211	&14.356	 &0.051	&14.371	 &0.025	&14.304	 &0.026	&14.224	 &0.034	&14.570	 &0.034\\
59,321.261	&14.486	 &0.052	&14.304	 &0.029	&14.241	 &0.028	&14.206	 &0.043	&14.591	 &0.033\\
59,322.201	&14.359	 &0.072	&14.337	 &0.033	&14.259	 &0.027	&14.191	 &0.039	&14.622	 &0.037\\
59,323.230	&14.473	 &0.045	&14.269	 &0.023	&14.244	 &0.019	&14.204	 &0.028	&14.611	 &0.029\\
59,324.163	&14.533	 &0.065	&14.351	 &0.030	&14.258	 &0.032	&14.179	 &0.048	&14.684	 &0.043\\
59,325.280	&...     &...	&...      &...	&...	 &...	&14.226	 &0.053	&...	 &...  \\
59,326.207	&14.617	 &0.094	&14.293	 &0.032	&14.287	 &0.029	&14.235	 &0.049	&14.668	 &0.038\\
59,327.174	&14.679	 &0.072	&14.324	 &0.039	&14.416	 &0.050	&14.282	 &0.047	&14.759	 &0.078\\
59,332.197	&14.811	 &0.088	&14.503	 &0.042	&14.598	 &0.028	&14.525	 &0.048	&14.988	 &0.037\\
59,333.224	&14.937	 &0.054	&14.595	 &0.023	&14.673	 &0.020	&14.620	 &0.032	&15.046	 &0.031\\
59,334.241	&15.068	 &0.093	&14.644	 &0.046	&14.709	 &0.042	&14.615	 &0.033	&15.050	 &0.040\\
59,337.248	&15.408	 &0.058	&14.833	 &0.027	&14.860	 &0.030	&14.797	 &0.041	&15.091	 &0.084\\
59,338.176	&15.502	 &0.068	&14.906	 &0.031	&14.914	 &0.032	&14.857	 &0.051	&15.098	 &0.038\\
59,339.200	&15.567	 &0.042	&14.898	 &0.024	&14.937	 &0.028	&14.889	 &0.048	&15.098	 &0.033\\
59,340.177	&15.846	 &0.078	&14.986	 &0.033	&14.978	 &0.033	&14.908	 &0.056	&15.042	 &0.044\\
59,342.170	&15.997	 &0.116	&15.013	 &0.038	&14.879	 &0.035	&14.931	 &0.066	&14.945	 &0.042\\
59,343.170	&16.115	 &0.119	&15.078	 &0.041	&14.962	 &0.033	&14.968	 &0.054	&14.942	 &0.042\\
59,344.223	&16.243	 &0.064	&15.169	 &0.034	&14.972	 &0.032	&14.997	 &0.044	&14.921	 &0.042\\
59,345.194	&16.222	 &0.075	&15.162	 &0.036	&14.993	 &0.032	&...	 &...	&14.900	 &0.041\\
59,346.184	&16.362	 &0.092	&15.244	 &0.038	&15.038	 &0.035	&15.046	 &0.052	&14.877	 &0.043\\
59,347.212	&16.498	 &0.054	&15.281	 &0.025	&15.007	 &0.023	&15.062	 &0.035	&14.870	 &0.027\\
59,348.229	&16.676	 &0.060	&15.352	 &0.033	&15.076	 &0.032	&15.091	 &0.043	&14.873	 &0.040\\
59,349.188	&16.909	 &0.166	&15.287	 &0.053	&15.033	 &0.041	&15.109	 &0.068	&14.862	 &0.056\\
59,350.243	&16.815	 &0.075	&15.442	 &0.032	&15.123	 &0.028	&15.162	 &0.033	&14.897	 &0.033\\
59,351.241	&16.883	 &0.072	&15.512	 &0.029	&15.146	 &0.024	&15.189	 &0.022	&14.871	 &0.029\\
59,355.207	&17.182	 &0.137	&15.752	 &0.044	&15.359	 &0.038	&15.466	 &0.067	&15.078	 &0.046\\
59,356.247	&17.323	 &0.112	&15.867	 &0.047	&15.491	 &0.039	&15.512	 &0.043	&15.152	 &0.052\\
59,357.180	&17.621	 &0.165	&15.898	 &0.066	&15.524	 &0.051	&15.617	 &0.060	&15.259	 &0.065\\
59,358.181	&...	 &...	&...	 &...	&...	 &...	&15.617	 &0.050	&...	 &...  \\
59,360.208	&17.548	 &0.119	&16.105	 &0.053	&15.688	 &0.045	&15.750	 &0.040	&15.419	 &0.054\\
59,361.220	&17.615	 &0.105	&16.083	 &0.043	&15.804	 &0.043	&15.804	 &0.048	&15.495	 &0.043\\
59,363.230	&17.613	 &0.070	&16.134	 &0.034	&15.854	 &0.034	&15.875	 &0.033	&15.610	 &0.045\\
59,364.205	&17.559	 &0.101	&16.194	 &0.036	&15.890	 &0.038	&15.934	 &0.048	&15.660	 &0.042\\
59,366.223	&17.572	 &0.099	&16.285	 &0.045	&15.984	 &0.046	&16.044	 &0.056	&15.809	 &0.056\\
59,367.245	&17.529	 &0.096	&16.185	 &0.038	&15.991	 &0.037	&16.047	 &0.079	&15.796	 &0.047\\
59,368.213	&17.560	 &0.098	&16.322	 &0.042	&16.096	 &0.042	&16.107	 &0.073	&15.945	 &0.076\\
59,369.207	&17.708	 &0.104	&16.299	 &0.047	&16.094	 &0.047	&16.147	 &0.068	&15.924	 &0.058\\
59,371.209	&17.660	 &0.098	&16.401	 &0.047	&16.177	 &0.061	&16.208	 &0.051	&16.014	 &0.063\\
59,377.220	&17.622	 &0.106	&16.497	 &0.049	&16.290	 &0.051	&16.345	 &0.100	&16.210	 &0.071\\
\enddata
\label{tab:kast}
\end{deluxetable*}

\startlongtable
\begin{deluxetable*}{lccc}
\tablecolumns{3} 
\tablewidth{50pc} 
\tabletypesize{\scriptsize}
\tablecaption{Unfiltered optical photometry of SN 2021hpr taken with Itagaki
Astronomical Observatory 0.35\,m telescope}
\tablehead{\colhead{MJD} &\colhead{Epoch$^a$} &\colhead{mag}}
\startdata
59,306.448&-15.66&17.7\\
59,307.457&-14.65&17.1\\
59,307.458&-14.65&17.2\\
59,309.525&-12.58&16.1\\
59,309.525&-12.58&16.0\\
59,310.525&-11.58&15.7\\
59,310.525&-11.58&15.6\\
59,311.608&-10.50&15.3\\
59,311.608&-10.50&15.4\\
59,312.609&-9.50&15.1\\
59,312.609&-9.50&15.1\\
59,313.608&-8.50&14.8\\
59,313.608&-8.50&14.8\\
59,314.612&-7.49&14.7\\
59,314.612&-7.49&14.7\\
59,315.603&-6.50&14.4\\
59,315.603&-6.50&14.4\\
59,318.616&-3.49&14.2\\
59,318.617&-3.49&14.2\\
\enddata
\tablenotetext{a}{Relative to the epoch of $B$-band maximum brightness (MJD = 59,322.11).}
\label{tab:lc_clear}
\end{deluxetable*}

\bibliography{PASPsample631}{}
\bibliographystyle{aasjournal}



\end{document}